\documentclass[twocolumn]{aastex61}

\newcommand{\prems}{pre-MS}
\newcommand{\zams}{ZAMS}

\newcommand{\vk}{$V - K_s$}
\newcommand{\vt}{$V_\textrm{\scriptsize{T}}$}
\newcommand{\vj}{$V_\textrm{\scriptsize{J}}$}
\newcommand{\vjk}{$V_\textrm{\scriptsize{APASS}} - K_s$}
\newcommand{\vks}{$V-K_s$}
\newcommand{\teff}{$T_{\textrm{\scriptsize{eff}}}$}
\newcommand{\prot}{$P_{\textrm{\scriptsize{rot}}}$}
\newcommand{\Msun}{\hbox{$\mathcal{M}^{\rm N}_{\odot}$}}
\newcommand{\logm}{$\log(M/\mathcal{M}^{\rm N}_{\odot})$}

\received{}
\revised{}
\accepted{June 7, 2017}
\submitjournal{ApJ}

\shorttitle{Angular Momentum Evolution of Young Stars in the Nearby
  Scorpius-Centaurus OB Association}
\shortauthors{Mellon et al.}

\begin{document}

\title{Angular Momentum Evolution of Young Stars\\ in the Nearby
  Scorpius-Centaurus OB Association}

\correspondingauthor{Samuel N. Mellon}
\email{smellon@ur.rochester.edu}
%\authorcomment1{ApJ, in press}

\author[0000-0003-3405-2864]{Samuel N. Mellon}
\affiliation{Department of Physics \& Astronomy, University of Rochester,
  Rochester, NY 14627, USA}
\affiliation{Department of Physics, Westminster College, New
  Wilmington, PA 16172, USA}

\author[0000-0003-2008-1488]{Eric E. Mamajek}
\affiliation{Jet Propulsion Laboratory, California Institute of Technology, M/S
  321-100, 4800 Oak Grove Drive, Pasadena, CA 91109, USA}
\affiliation{Department of Physics \& Astronomy, University of
  Rochester, Rochester, NY 14627, USA}
  
\author[0000-0002-7885-8475]{Thomas E. Oberst}
\affiliation{Department of Physics, Westminster College, New
  Wilmington, PA 16172, USA}
  
\author[0000-0002-7859-1504]{Mark J. Pecaut}
\affiliation{Department of Physics, Rockhurst University, Kansas City,
  MO 64110, USA}

\begin{abstract}
We report the results of a study of archival SuperWASP light curves
for stars in Scorpius-Centaurus (Sco-Cen), the nearest OB association.
We use SuperWASP time-series photometry to extract rotation periods
for 189 candidate members of the Sco-Cen complex, and verify that 162
of those are members of the classic Sco-Cen subgroups of Upper
Scorpius (US), Upper Centaurus-Lupus (UCL), and Lower Centaurus-Crux
(LCC).
This study provides the first measurements of rotation periods
explicitly for large samples of pre-main sequence (pre-MS) stars
spanning the UCL and LCC subgroups.
Our final sample of 157 well-characterized pre-MS stars spans ages of
$\sim$ 10 -- 20 Myr, spectral types of $\sim$ F3 -- M0, and masses of $M
\simeq$ 0.3 -- 1.5 \Msun.
For this sample, we find a distribution of stellar rotation periods
with a median of $P_{\textrm{\footnotesize{rot}}} \simeq$ 2.4 days,
overall range of 0.2 $<$ $P_{\textrm{\footnotesize{rot}}}$ $<$ 8 days, and
a fairly well-defined mass-dependent upper envelope of rotation
periods.
This distribution of periods is consistent with recently developed
stellar angular momentum evolution models.
These data are significant because they represent an undersampled age
range and the number of measurable rotation periods is large compared
to recent studies of other regions.
We also search for new examples of eclipsing disk or ring systems
analogous to 1SWASP J140747.93-394542.6 (``J1407'', V1400 Cen), but
find none.
Our survey yielded five eclipsing binaries, but only one appears to be
physically associated with the Sco-Cen complex.
V2394 Oph is a heavily reddened (A$_V$ $\simeq$ 5 mag) massive
contact binary in the LDN 1689 cloud whose Gaia astrometry is clearly
consistent with kinematic membership with the Ophiuchus star-forming
region.
\end{abstract}

\keywords{
open clusters and associations: individual (Lower Centaurus-Crux,
Lupus, Ophiuchus, Upper Centaurus-Lupus, Upper Scorpius, Sco-Cen) ---
stars: binaries: eclipsing ---
stars: pre-main sequence ---
stars: rotation ---
stars: starspots
}

\section{Introduction \label{Sec:Intro}}

The Scorpius-Centaurus OB Association (Sco-Cen) is the nearest OB
association to the Sun \citep[d $\simeq$ 118 -- 145 pc;][]{deZeeuw99,
  Preibisch08}.
It contains the nearest {\it large} sample of 10 -- 20 Myr stars, making
it valuable for direct imaging of giant exoplanets and studies of disk
evolution.
The group is composed of three classically defined subgroups: Upper
Scorpius (US; median age $\simeq$ 11 Myr), Upper Centaurus-Lupus
(UCL; median age $\simeq$ 16 Myr), and Lower Centaurus-Crux (LCC; 
median age $\simeq$ 17 Myr) \citep{Pecaut12, Pecaut16}.
This grouping can be problematic, however, as the boundaries of the
subgroups are somewhat ill-defined and each group exhibits significant
substructure \citep{Preibisch08, Rizzuto12, Pecaut16}.
Most stars located in the three subgroups with masses of $<$ 2
\Msun\footnote{\Msun\, is the symbol for the nominal solar mass as
  defined by IAU Resolution 2015 B3 \citep{Prsa16}).}
are pre-main sequence (\prems), and some are still accreting from
protoplanetary disks \citep[e.g.][]{Luhman12, Pecaut16}.
Throughout this paper, we refer to the collection of the classic
subgroups US, UCL, and LCC as the Sco-Cen OB Association. We refer to
the ensemble of active and recent star-formation in the vicinity of
the Sco-Cen association as the {\it Sco-Cen complex}, including the
young associations in the Oph, Lup, CrA, and Cha molecular clouds, and
the smaller peripheral groups of $\sim$ 5 -- 10 Myr-old stars
\citep[$\epsilon$ Cha, $\eta$ Cha, and TW Hya;][]{Preibisch08}. These
regions represent a large-scale star-formation event that has been
occurring over the past $\sim$ 20 Myr, forming discrete subgroups of
batches of dozens to thousands of stars during that span. The three
classic subgroups likely represents ensembles of numerous smaller
star-formation events rather than monolithic bursts of star formation
\citep{Pecaut16}.

% SEC 1, PAR 2

The distribution of rotation periods for \prems\, stars in Sco-Cen can
provide useful constraints on stellar angular momentum evolution
models. The angular momentum evolution of young stars is governed by
several processes that work to increase or decrease the rotation speed
of the star.
During the \prems\, portion of a star's lifetime the gravitational
contraction lowers the star's moment of inertia, which can increase
the angular rotation speed as a consequence of angular momentum
conservation; magnetic “disk-locking” can also work against this
contraction \citep{Irwin11, Gallet13}.
Beyond the zero-age main sequence (\zams), the moment of inertia
changes very slowly and the star's angular momentum evolution is
dominated by braking via magnetized stellar winds and the transfer of
angular momentum between the interior and exterior layers of the star
causing a steady spin-down for the remainder of the star's life
\citep{Gallet15}.

% SEC 1, PAR 3

In recent years, rotation periods have been measured for hundreds of
stars over a wide range of masses \citep[e.g.][]{Hartman08, Hartman10,
  Messina10, Meibom11a, Meibom11b, Irwin11, Gallet13, Moraux13,
  Cargile14, Meibom15, Gallet15, Douglas16}.
These distributions of rotation periods as a function of stellar age
have enabled the development of angular momentum evolution models.
These models, which are used to estimate the ages of stars from the
earliest \prems\ through the end of the MS, take into account
disk-locking, gravitational contraction, stellar winds, and many other
factors \citep[e.g.][]{Meibom11a, Barnes10b, Reiners12, Bouvier14,
  Cargile14, Gallet15}.
Additional large surveys of rotation periods during the post-accretion
pre-MS phase can help constrain these angular momentum evolution
models as this age range is undersampled \citep[e.g.][]{Gallet15}.
%
%Sec1, Par 4

The work presented in this paper found rotation periods for 189 young
stars, 96 of which are newly measured periods (including an outlying K
giant star with a newly measured short activity period). 162 of these
stars belong to the three classic subgroups (US, UCL, LCC). 157 of the
Sco-Cen members had retrievable spectral types, which were used to
estimate the masses of each star. The stars with spectral types were
then plotted against current theoretical angular momentum evolution
models from \citet{Gallet15}. This study finds that these data are
consistent with what these models predict.

This study was also designed to discover and characterize new
circumsecondary eclipsing disk/ring systems like the one found around
J1407 \citep[V1400 Cen;][]{Mamajek12, Scott14, Kenworthy15} and new
examples of rare \prems\, eclipsing binary stars
\citep[e.g.][]{Morales-Calderon12, Kraus15}.
No new eclipsing disk systems were discovered.
Five candidate eclipsing binary systems were identified.
However, upon further scrutiny, only {\it one} appears to be
associated with Sco-Cen.

% SEC 1, PAR 5

This paper is organized as follows:
{\S 2} discusses the construction of our initial sample of Sco-Cen
candidates and details the data used in the survey;
{\S 3} details the use of periodograms and generation of phase-folded light curves;
{\S 4} provides a summary of the study's results, discusses the
rotational periods of the Sco-Cen members and how these rotational
periods help to understand the evolution of low-mass stars.

%%% SECTION 2: DATA 

\section{Data \label{Sec:Data}}

\subsection{Sco-Cen Sample \& Membership \label{Sec:Membership}} %%% SEC 2.1 

In the course of previous work on the membership and star formation
history of the Sco-Cen complex \citep[including both the OB subgroups
  and related young stellar object populations in the associated
  molecular clouds; see e.g. ][]{Preibisch08, Pecaut12, Pecaut16}, an
internal database was constructed of 5,551 candidate stellar members.
For the classic subgroups, candidate
members were drawn from the following studies:
\citet{Ardila00},
\citet{Blaauw46},
\citet{Dawson11},
\citet{deGeus89},
\citet{Hoogerwerf00},
\citet{Lodieu06},
\citet{Lodieu07},
\citet{Lodieu13},
\citet{Luhman12},
\citet{Mamajek02},
\citet{Martin04},
\citet{Pecaut12},
\citet{Pecaut16},
\citet{Preibisch08},
\citet{Rizzuto12},
\citet{Sartori03},
\citet{Slesnick06},
\citet{Song12},
\citet{Wichmann97},
and \citet{deZeeuw99}.
The quality of membership assignments in these studies is quite
heterogeneous.
Some were selected only by virtue of photometry, proper motions, and/or
X-ray emission.
Many were also vetted using parallaxes and proper motions from
  the first Gaia data release \citep{Gaia2016}.
Given the large number of candidate members, memberships were only
reassessed if a star passed several criteria summarized in {\S
  \ref{pgrams}}.

\subsection{Photometry \label{Sec:Photom}} %%% SEC 2.2 

\subsubsection{SuperWASP Photometry \label{Sec:Time_Photom}} %%% SEC 2.2.1

In order to estimate both long-term median magntiudes and rotational
periods (i.e. periodic variations in magnitudes due to starspot
rotation) for Sco-Cen stars, these 5,551 candidate members were
cross-referenced with the archival single-band time-series photometric
data catalogue from the Super Wide Angle Search for Planets
(SuperWASP). SuperWASP consists of two robotic observatories in La
Palma, Spain, and Sutherland, South Africa. Each observatory has a
bank of eight wide-angle cameras that collectively provide a 490
deg$^2$ field of view (FOV) per pointing at 13\arcsec\,pix$^{-1}$ scale
within magnitude range $8 < V < 13$. The observatories have been
operating simultaneously and year-round since 2004 to collect $V$-band
photometry over most of the sky with a single-position cadence of
approximately 10 minutes \citep{Pollacco06, Butters10, SmithA14}.

% SEC 2.2.1, PAR 2

The first and only SuperWASP public data release includes data
collected between 2004 and 2008. It is available for download via the
NASA Exoplanet Archive operated at the NASA Exoplanet Science
Institute\footnote{http://exoplanetarchive.ipac.caltech.edu/}. The
data were processed via the SuperWASP pipeline and post-pipeline
analysis including astrometric calibration, aperture photometry, and
photometric calibration. The data are provided in *.FITS and *.tbl
formats, the latter containing observation timestamps (HJD),
magnitudes calibrated to the Tycho-2 \vt\, system, and their
uncertainties \citep{Pollacco06, Butters10}.  SuperWASP fields in the
Sco-Cen region were covered over three $\sim$ 100 day seasons between
2006 and 2008. Of the 5,551 candidate Sco-Cen members, 1,689 of them
were found to have SuperWASP counterparts.

% SEC 2.2.1, PAR 3

A data reduction and periodogram analysis pipeline was used to process
SuperWASP time series photometry (described in
\S\ref{Sec:Periodogram}) for each of the 1,689 candidate stars. Table
\ref{Tab:Photometry} contains the raw results from 1689 light curves,
which includes the strongest periods (excluding obvious aliases) and
fitted amplitudes for these periods from each season.

\begin{figure*}[hbt]
\centering
\includegraphics[width=0.9\textwidth]{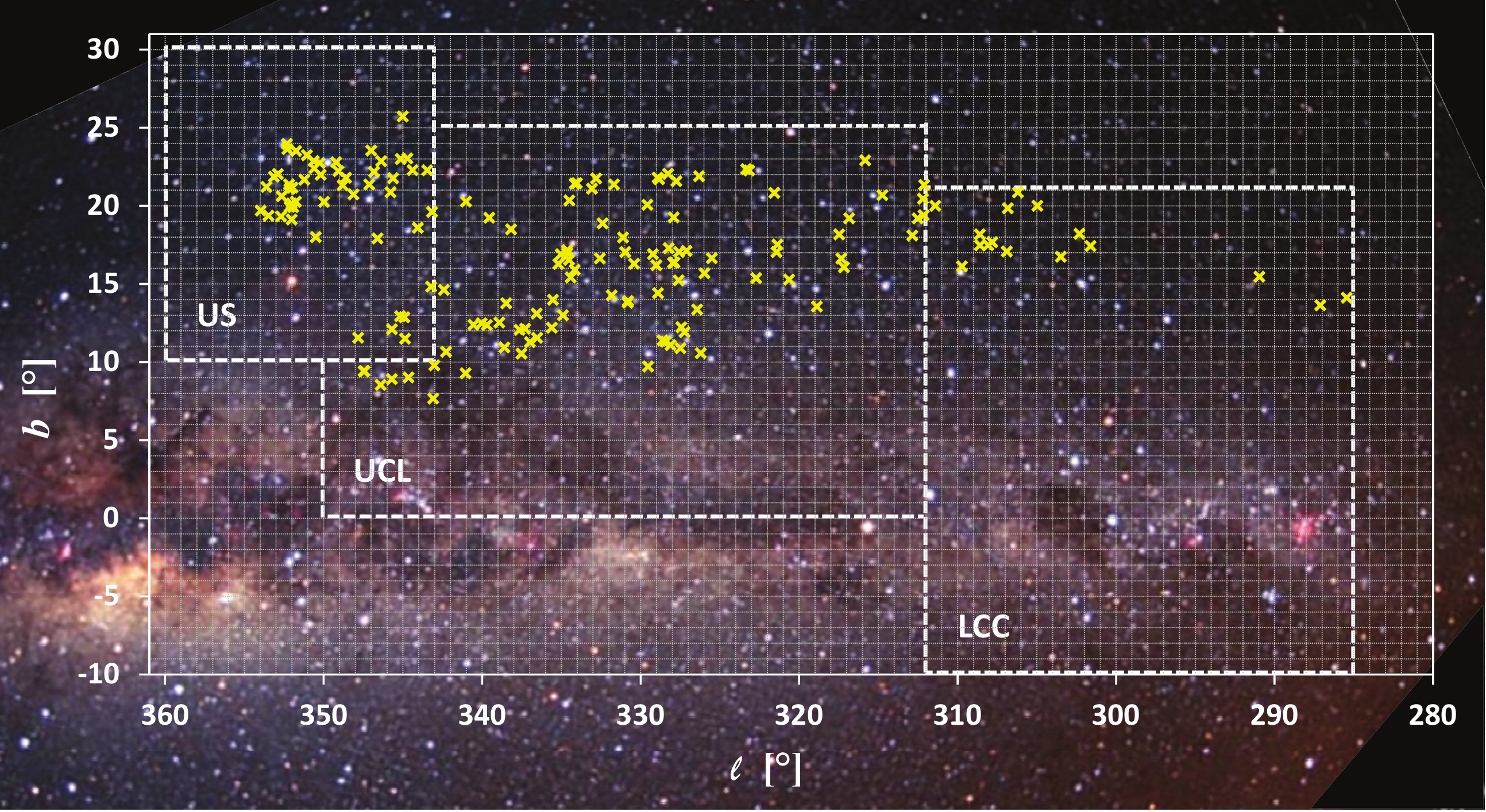}
\caption{The distribution of Sco-Cen members with rotation periods
  plotted in Galactic Coordinates (yellow crosses).  The background is
  a wide field optical image of the Sco-Cen region taken by Fred
  Espenak (NASA GSFC; \textit{www.mreclipse.com}). The classic Sco-Cen
  subgroup boundaries defined by \citet{deZeeuw99} are plotted as
  dashed white lines. \label{Fig:lb}}
\end{figure*}

Only 189 of them were found to have rotation periods that were
consistent throughout at least two of the 2006 -- 2008 seasons. Of
those 189 stars, 162 of them were confirmed as members of classic
subgroups (plotted in Figure \ref{Fig:lb}, see
\S\ref{Sec:Multi}). Table \ref{Tab:Results} contains the reduced
period and stellar information for these 162 stars. This sample is
further trimmed to 157 stars for aspects of the analysis requiring
spectral types since seven of the stars do not have a measured
spectral type.

The remaining 27 stars belong to associated younger star-forming
regions in the Sco-Cen complex, namely the Lup, Oph, and CrA regions
(and two foreground members of the TW Hya association and a background
Li-rich K giant). These 27 stars were not used in the subsequent
analysis as their numbers are small (the effects of extinction are
likely greatly reducing the coverage of members of these star-forming
regions in SuperWASP catalog). These stars are compiled in Table
\ref{Tab:Other_Stars} including their statistically significant
periods.

% FIGURE 1: MAP OF SCO-CEN

%%% SEC 2.2.2

\subsubsection{Multi-Band Photometry \label{Sec:Multi}} %%% SEC 2.2.2 

Due to the large pixel size of the SuperWASP survey (13.7''
pixel$^{-1}$), the median \vt\ magnitude may represent the unresolved
light from multiple stars, which can result in spurious estimates of
colors, reddening, and extinction (see {\S
  \ref{Sec:Intrinsic}}). Thus, in addition to SuperWASP single-band
time-series photometry, multi-band single instance (non-time-series)
photometry from the archival databases of the Two Micron All-Sky
Survey \citep[2MASS;][]{Cutri032MASS,Skrutskie06}, the fourth United
States Naval Observatory CCD Astrograph Catalog,
\citep[APASS;][]{Henden12}, and the Yale/San Juan Southern Proper
Motion Catalog 4 \citep[SPM4;][]{Girard11} were used. These data were
employed for three purposes: (1) to convert SuperWASP's Tycho-2 \vt\,
magnitudes to Johnson \vj\, magnitudes; (2) to help verify the
membership of our Sco-Cen candidates; and (3) to obtain accurate
$V$-band magnitudes for estimates of observed colors, estimated
reddenings, and HR diagram placement for the target stars.

%%% SEC 2.2.2, PAR 3

To convert SuperWASP's Tycho-2 \vt\, magnitudes to the Johnson \vj\,
system, two linear trends were fit to the Tycho-2 \citep{Hog00} and
2MASS \citep{Skrutskie06} photometry for nearby ($d$ $<$ 75 pc) {\it
  Hipparcos} stars whose absolute magnitudes were within 1 mag of the
main sequence, and whose photometric errors were $<$ 0.03 mag in the
relevant bands:

\begin{equation}
\begin{footnotesize}
V_{\textrm{\scriptsize{J}}} - V_{\textrm{\scriptsize{T}}} = -0.095 - 0.062(V_{\textrm{\scriptsize{T}}} - J - 1.631)
\end{footnotesize}
\end{equation}

\begin{equation}
\begin{footnotesize}
V_{\textrm{\scriptsize{J}}} - V_{\textrm{\scriptsize{T}}} = -0.083 - 0.049(V_{\textrm{\scriptsize{T}}} - H - 1.791)
\end{footnotesize}
\end{equation}

\noindent The uncertainties in the zero-points are 0.001 mag, and
uncertainties in the slopes are 0.002, while the rms dispersions in
the fits are 0.01 mag. The fits are well-constrained over the color
ranges 0 $<$ (\vt\ - $J$) $<$ 3.6 and 0 $<$ (\vt\ -
$H$) $<$ 4.2.

%%% SEC 2.2.2, PAR 4

While cross-referencing the SuperWASP and 2MASS data, we found that,
for approximately one-third of the stars in our sample, the best
spatial matches had poor brightness matches. The worst offenders
revealed themselves as unphysical outliers on a color-magnitude
diagram (\vj\, vs. \vjk). These cases were found to each be caused by
two 2MASS targets of significant ($\gtrsim$ 0.75 magnitude) brightness
difference existing in close spatial proximity to a single SuperWASP
target. Each was corrected by simply selecting the 2MASS counterpart
as the one of (obvious) comparable brightness to the SuperWASP target.

%%% SEC 2.2.2, PAR 5

\begin{figure}[hbtp] 
\centering
\includegraphics[width=0.5\textwidth]{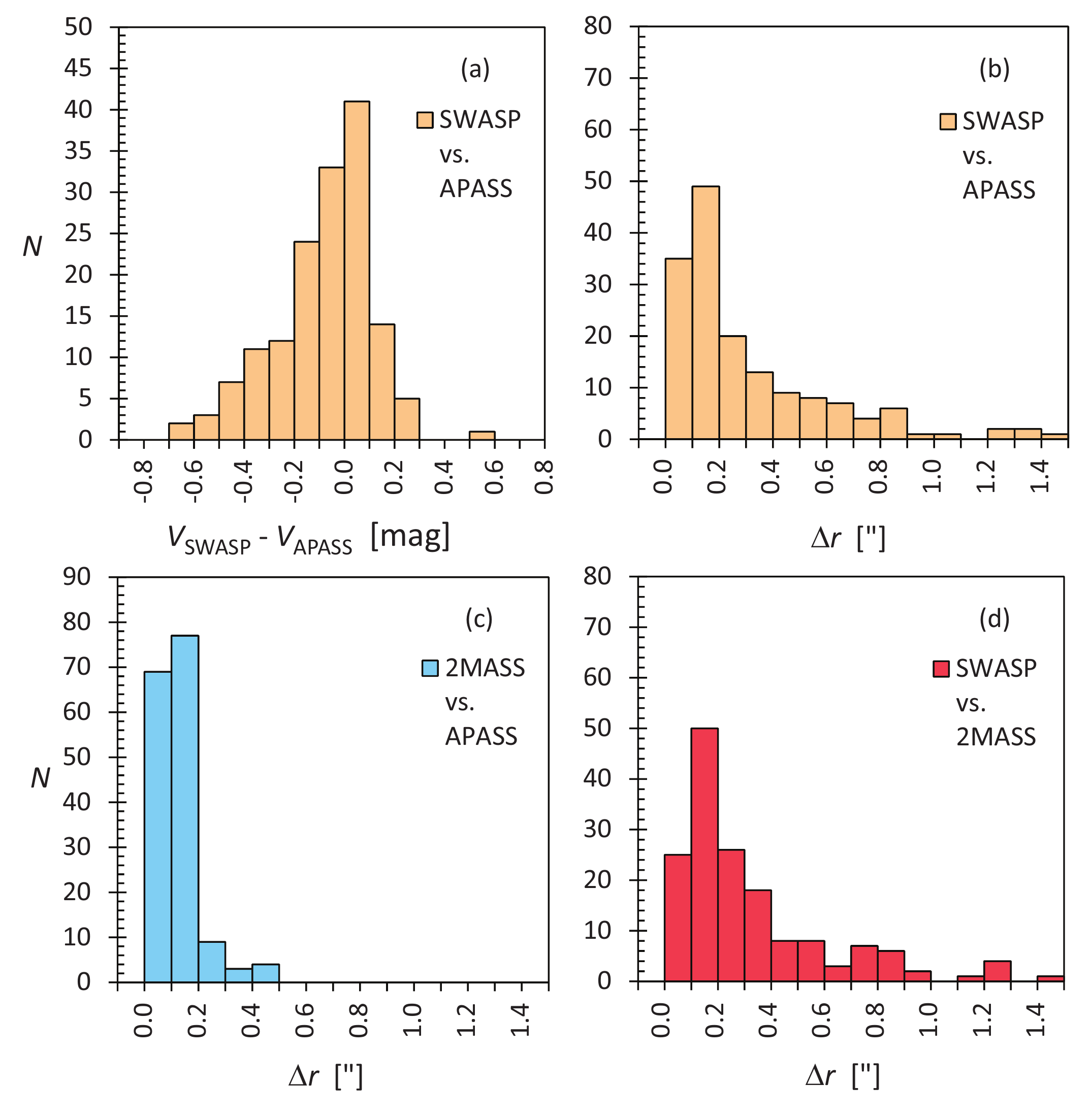}
\caption{Select distributions of spatial and brightness differences
  between the SuperWASP, APASS, and 2MASS data sets for the final 162
  star sample. In Panel (a) the brightnesses for both surveys are in
  Johnson $V$ magnitudes. The remaining panels show the J2000
  positional differences. \label{Fig:Histograms}} 
\end{figure}

Many smaller but significant ($0.2 \lesssim \Delta{V} \lesssim 0.75$
magnitudes) brightness differences remained in cases with no spatial
degeneracy of 2MASS counterparts. This prompted us to compare the
SuperWASP and 2MASS brightnesses and positions with those of the APASS
and SPM4 surveys as an additional check. The 2MASS, APASS, and SPM4
brightnesses were found to be in excellent agreement: e.g. $(\Delta
V)_{\rm average}$ $\simeq$ 0.01 magnitudes for the APASS and SPM4
catalogs. However, the SuperWASP brightnesses were found to have much
poorer agreement: e.g. $(\Delta V)_{\rm average}$ $\simeq$ 0.1 mag
when comparing the converted SuperWASP $V$ magnitudes to either of the
APASS or SPM4 catalogs, a factor of 10 worse. {Figure
  \ref{Fig:Histograms}} panel (a) shows the distribution of brightness
differences between SuperWASP and APASS. The sizable skewing of
SuperWASP data to brighter magnitudes can be attributed to blending
due to SuperWASP's large pixel scale. Panels (b), (c) and (d) show the
J2000 positional differences between the four surveys. While these
differences are all small at ($\Delta r)_{\rm average}$ $<$
0.6\arcsec\, or better, the APASS positions agree best with those of
2MASS at ($\Delta r$)$_{\rm average}$ = 0.12\arcsec.

%%% SEC 2.2.2, PAR 6

\begin{figure}[h] 
\centering
\includegraphics[width=0.5\textwidth]{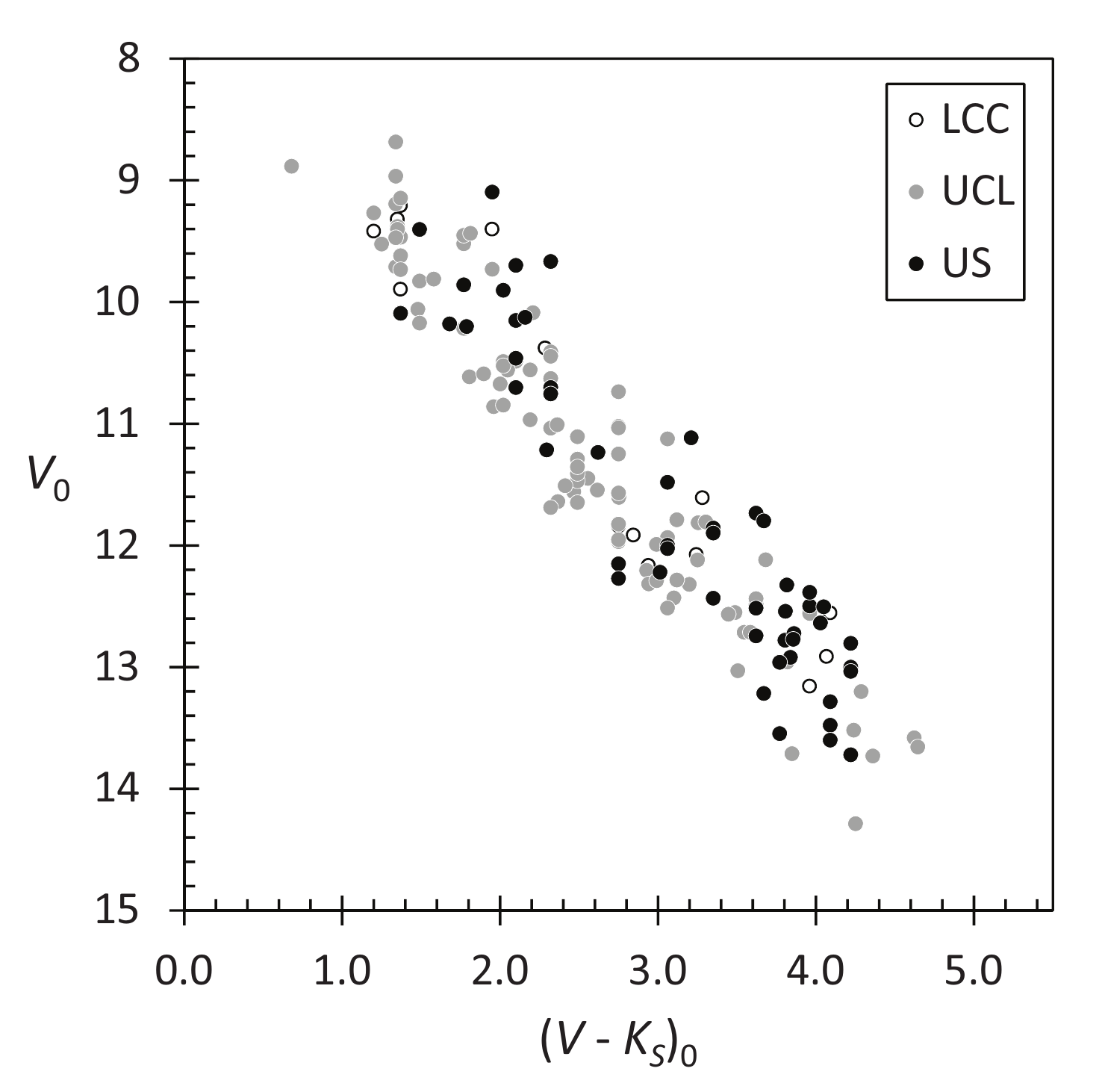}
\caption{Dereddened color-magnitude diagram for the final sample of
  162 stars in the classic Sco-Cen subgroups of US (filled black
  circles), UCL (filled gray circles), and LCC (open circles). The
  $V$-band brightness on both axes is APASS Johnson $V$, and the $K_S$
  value is from 2MASS. \label{Fig:Color-Mag}}
\end{figure}

APASS Johnson $V$-band magnitudes were used instead of the converted
SuperWASP $V$ magnitudes for all other aspects of our study
(e.g. colors, reddening, HR diagram analysis). All of this gave us the
highest confidence in the APASS $V$ magnitudes and the lowest
confidence in the SuperWASP $V$ magnitudes. SPM4 Johnson $V$
magnitudes were adopted only for the five stars in our 162-star sample
for which APASS $V$ magnitudes are not available Finally, we computed
\vks\, colors for the stars in our sample using the adopted $V$
magnitudes (from APASS or SPM4) and 2MASS $K_S$ magnitudes. Figure
\ref{Fig:Color-Mag} shows a de-reddened color-magnitude diagram for
the final sample of 162 members of the three Sco-Cen subgroups. The
treatment used for interstellar reddening and extinction is discussed
later in \S\ref{Sec:Intrinsic}.

%%% FIGURE 2: MAGNITUDE AND POSITION DIFFERENCES

%%% FIGURE 3: V-K COLOR-MAG DIAGRAM

%%% SECTION 3

\section{Analysis \label{Sec:Analysis}}

\subsection{Time-Series Analysis \label{Sec:Periodogram}}

\subsubsection{Data Reduction \label{reduction}} %%% SEC 3.1.1

\begin{figure}[hbtp]
\includegraphics[width = 0.5\textwidth]{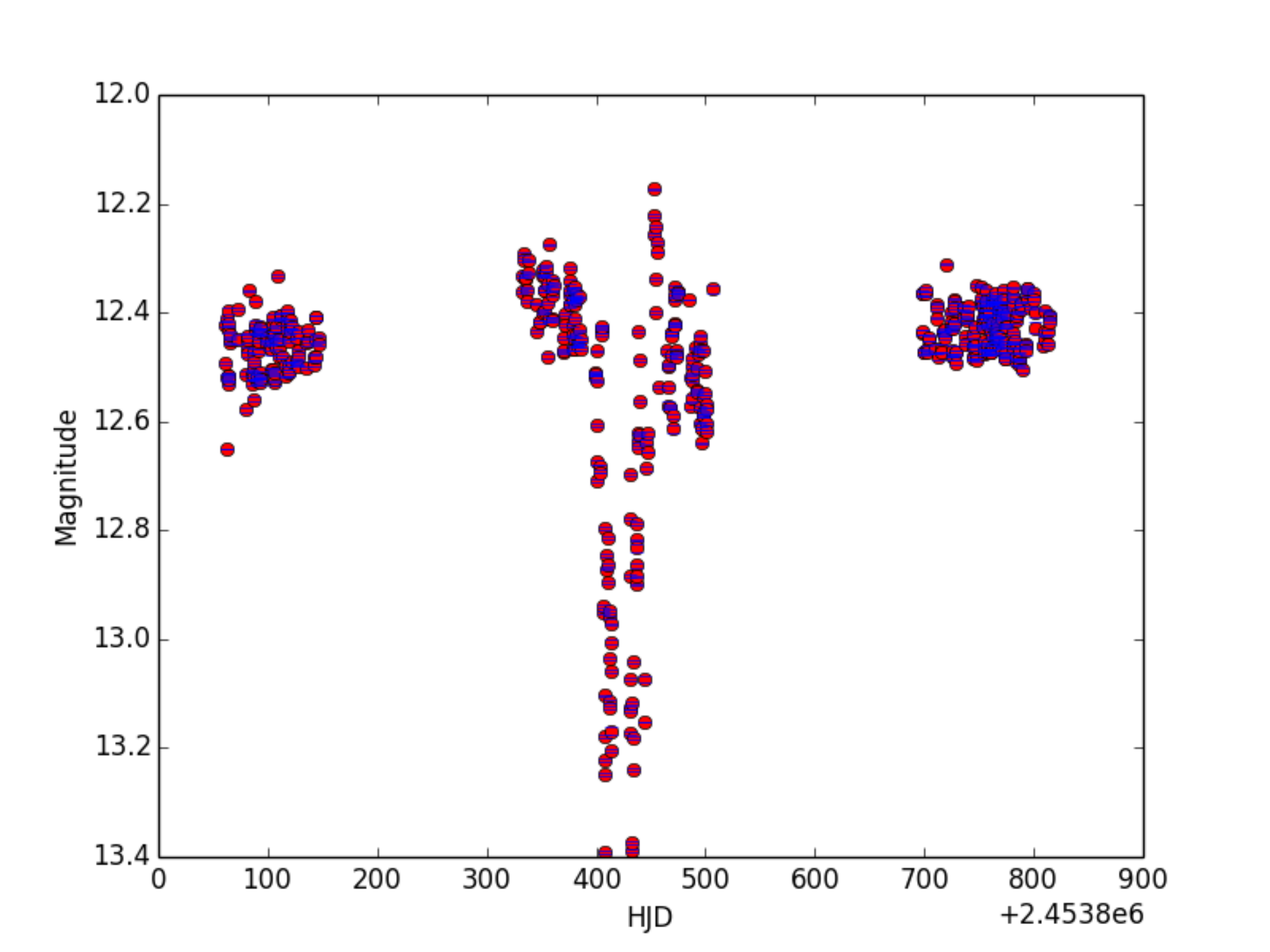}
\caption{The reduced SuperWASP light curve for V1400 Cen (``J1407'')
  over three observing seasons. Error bars (blue) are smaller than the
  data points themselves. The transit reported in \citet{Mamajek12} is
  clearly detected. \label{J1407_LC}}
\end{figure}

SuperWASP photometric data requires additional data reduction beyond
its own pipelines \citep{Cameron06}.
Reported photometric errors, median binning, and a 3$\sigma$ clip were
used to further reduce the data.
Data points with reported photometric errors of $>$ 0.1 mag were
removed.
Each data set was median combined and the standard error of the mean
for each binned point was recorded; the low-period limit of 0.1 days
was used because stars are not expected to have rotation periods
shorter than this due to instability \citep{Hartman10}.
Bins with fewer than 3 points were removed.
A 3$\sigma$ clip was applied to the binned data to remove any
remaining spurious points.
The reduced data was separated into three individual
$\sim$ 100 day time frames (seasons).
Finally, a plot of the full light curve, with error bars, was
generated ({Figure \ref{J1407_LC}}).

%%%FIGURE 4: REDUCED LC EXAMPLE

\subsubsection{Periodograms} %%% SEC 3.1.2
\label{pgrams}

A custom periodogram analysis pipeline was written using the
Python\footnote{http://python.org} language. Modules from the
SciPy\footnote{http://www.scipy.org/} stack: scipy, numpy, matplotlib,
and pandas \citep[][]{Scipy, Numpy, Matplotlib, Pandas} were used for
reducing and plotting data (with error bars) ({\S \ref{reduction}}),
performing Lomb-Scargle periodograms, and generating phase-folded
light curves with fitted amplitudes.

The Lomb-Scargle (LS) periodogram \citep{Press92, Scargle82} is a
powerful tool for extracting periods from unevenly sampled time series
photometry data sets \citep[e.g.][]{Hartman08, Hartman10, Messina10}.
The LS periodogram routine from the Python {\it scipy} module was
applied to each observing season for each star.
The period range spans from 0.1 days (see {\S \ref{reduction}}) to
$\tau$ days with a period step size of 0.15 days where $\tau$ is
half the length of time for the data collected in a particular season.
Periods corresponding to half the time length of the data set ($\tau$)
were searched to ensure that all possible long term periods in the
data set can be detected.
The period stepsize provided a resolution in the periodogram fine
enough to detect strong periods accurately over a large period span
without being overly computationally expensive.
After sorting data points into their individual seasons, the routine
returned the normalized values, which were then plotted versus
frequency.

We report the strongest period from each star’s periodogram in {Table
  \ref{Tab:Photometry}}. The LS periodogram routine alone does not
  estimate false positive periods, so the method of \citet{Cargile14}
  was employed to calculate false alarm probabilities (FAPs). In the
  interest of computational time, FAPs were only calculated for our
  final 162 star sample ({Table \ref{Tab:Results}}) and the 27
  ``other" stars ({Table \ref{Tab:Other_Stars}}).

\begin{figure*}[h]
\includegraphics[width=.33\textwidth]{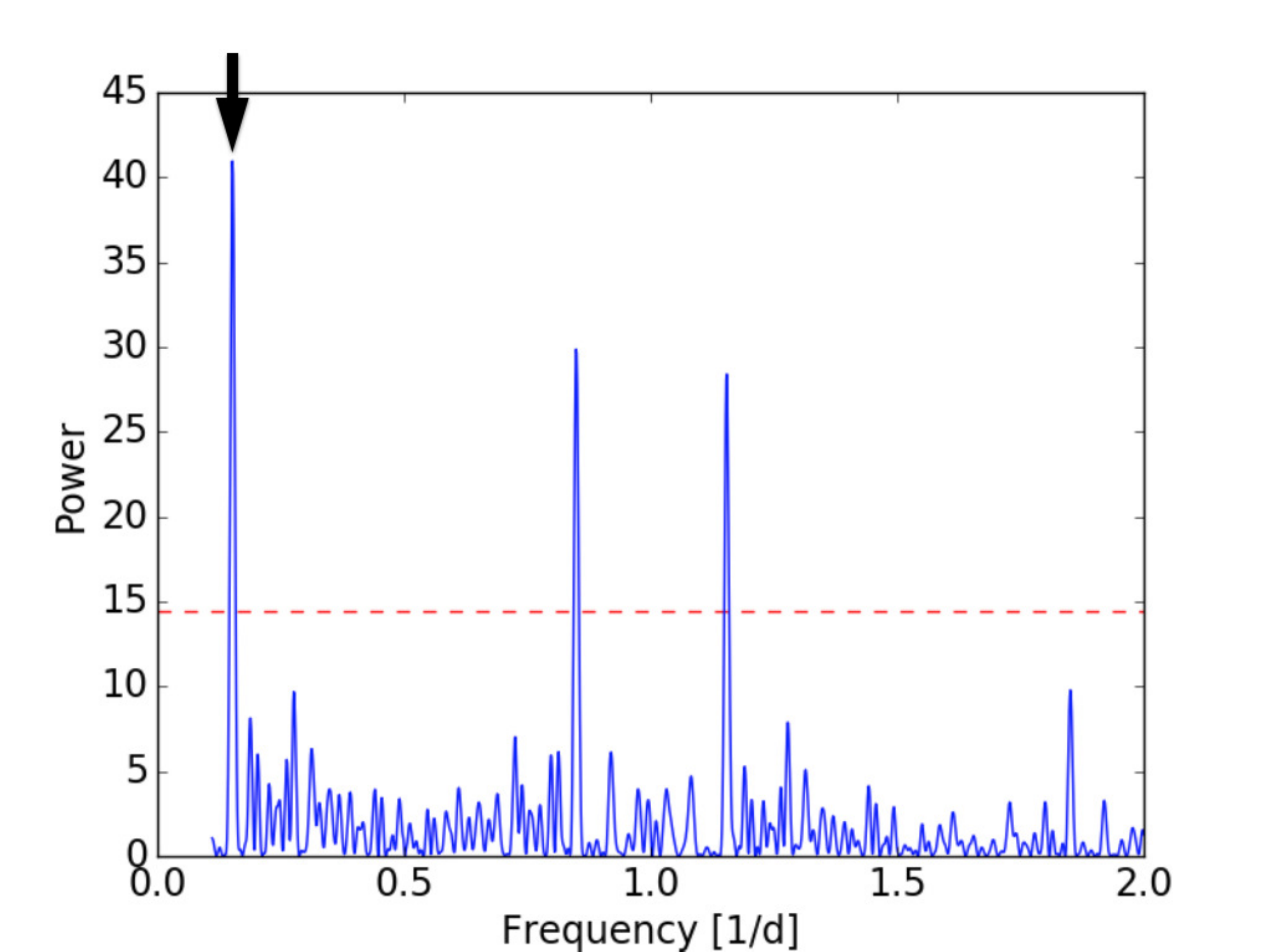}
\includegraphics[width=.33\textwidth]{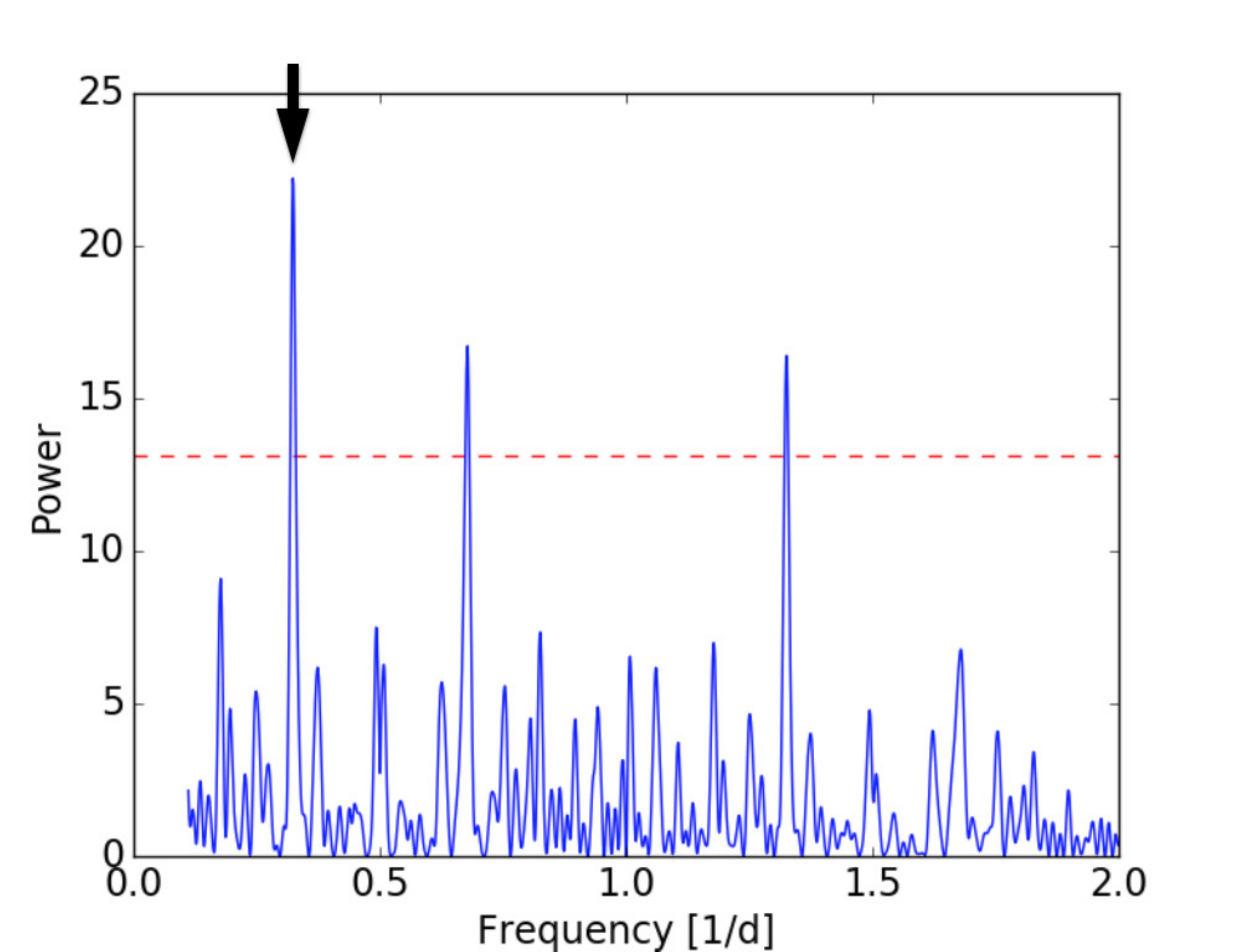}
\includegraphics[width=.33\textwidth]{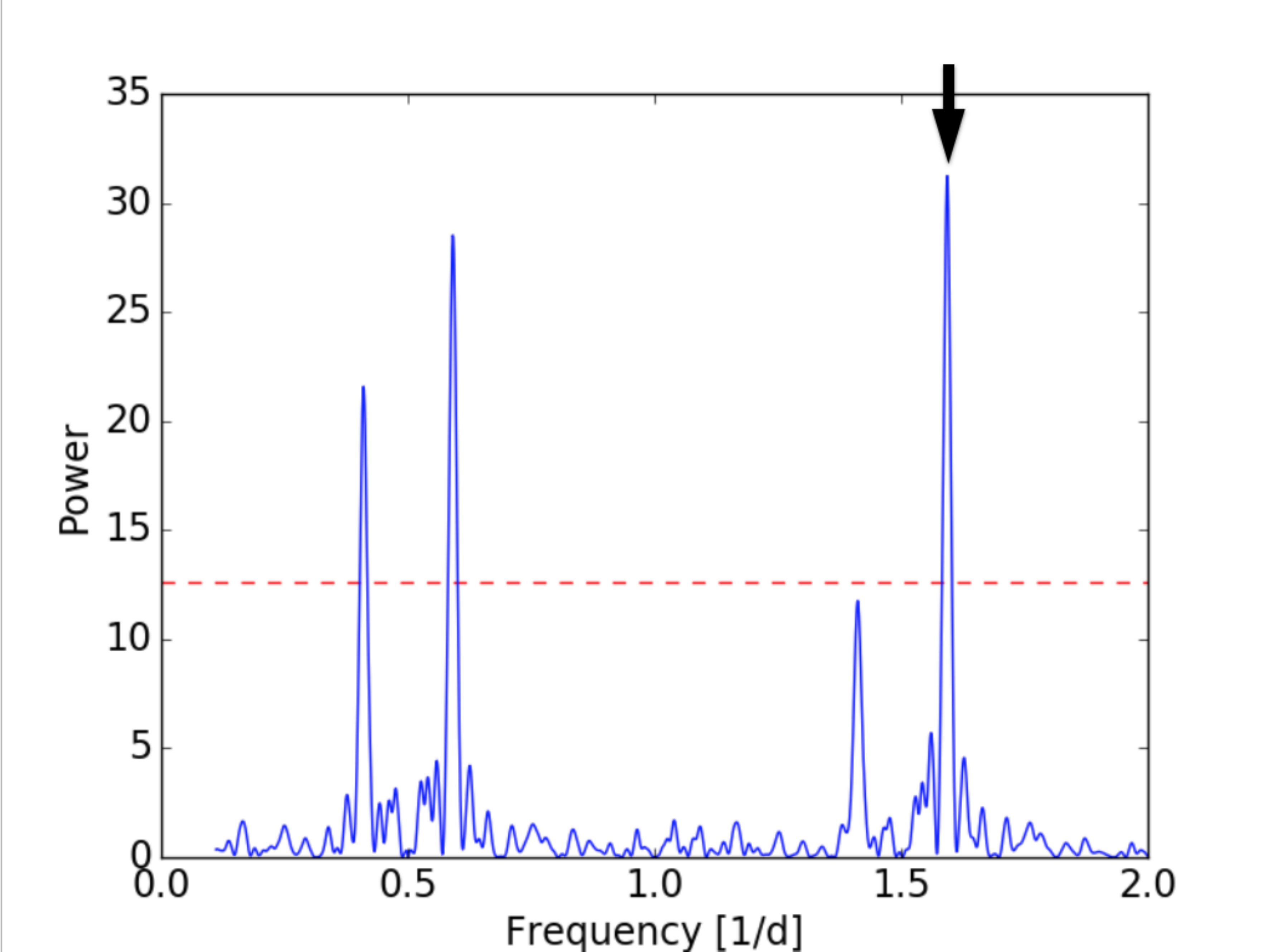}
\includegraphics[width=.33\textwidth]{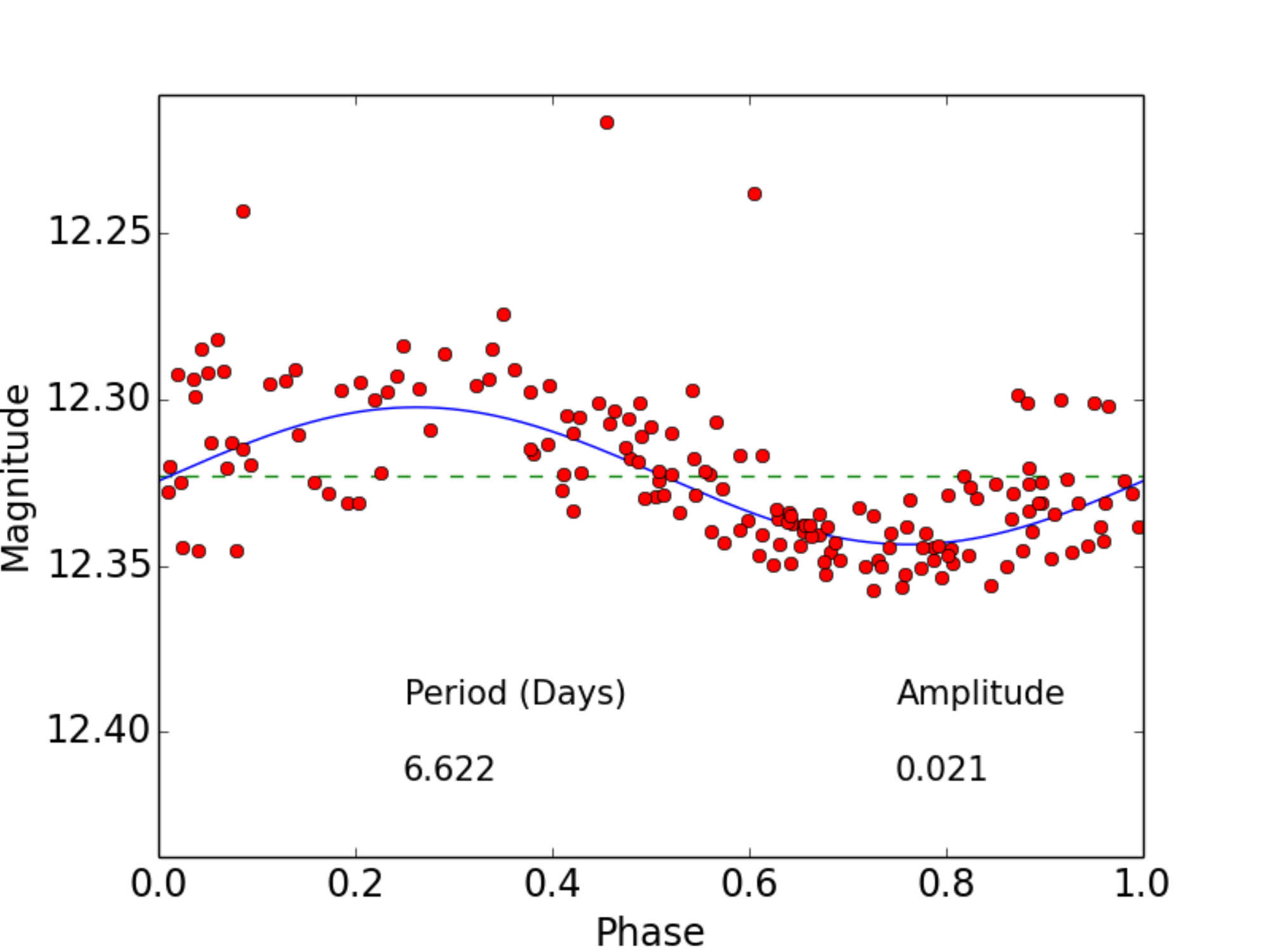}
\includegraphics[width=.33\textwidth]{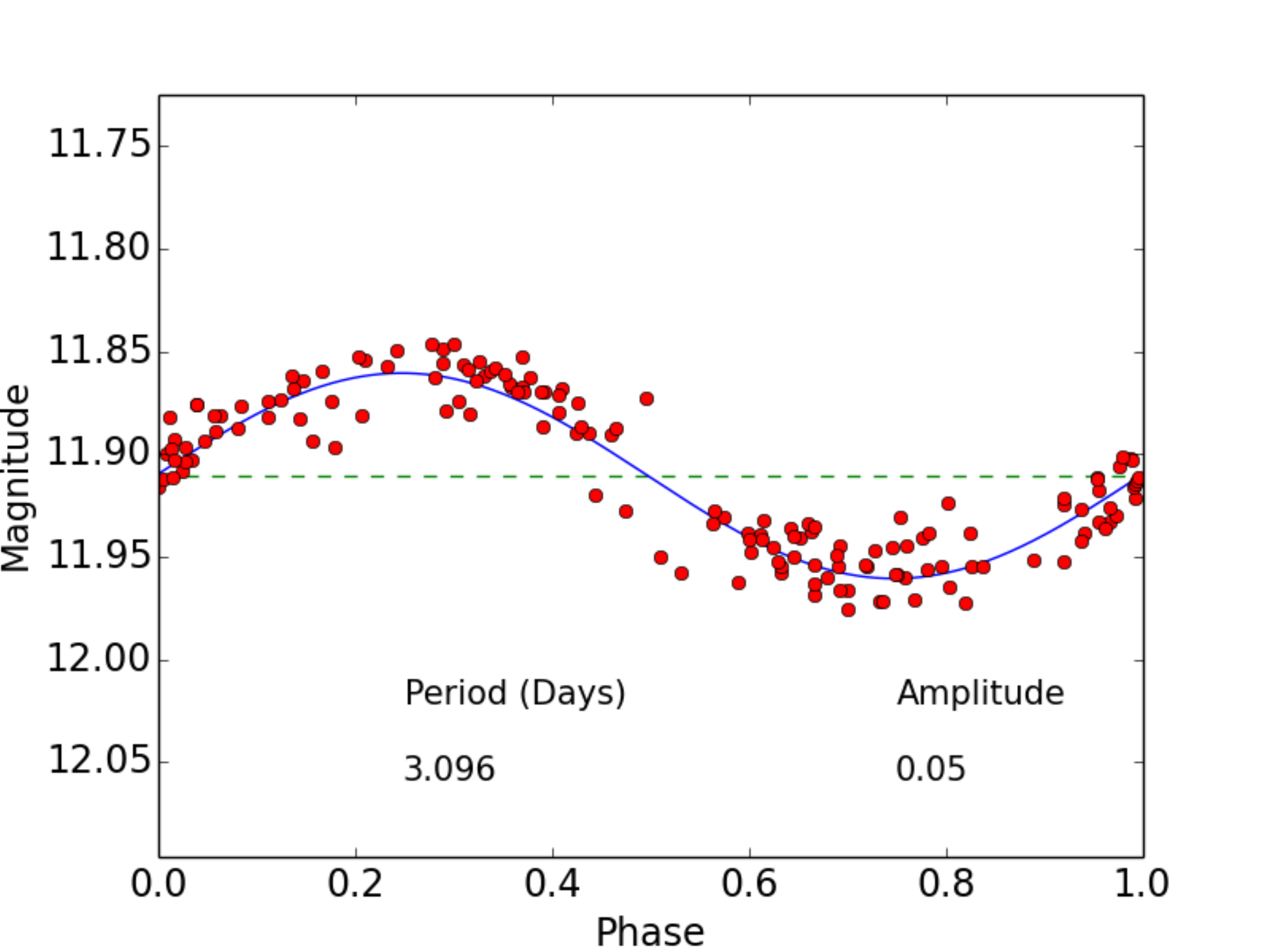}
\includegraphics[width=.33\textwidth]{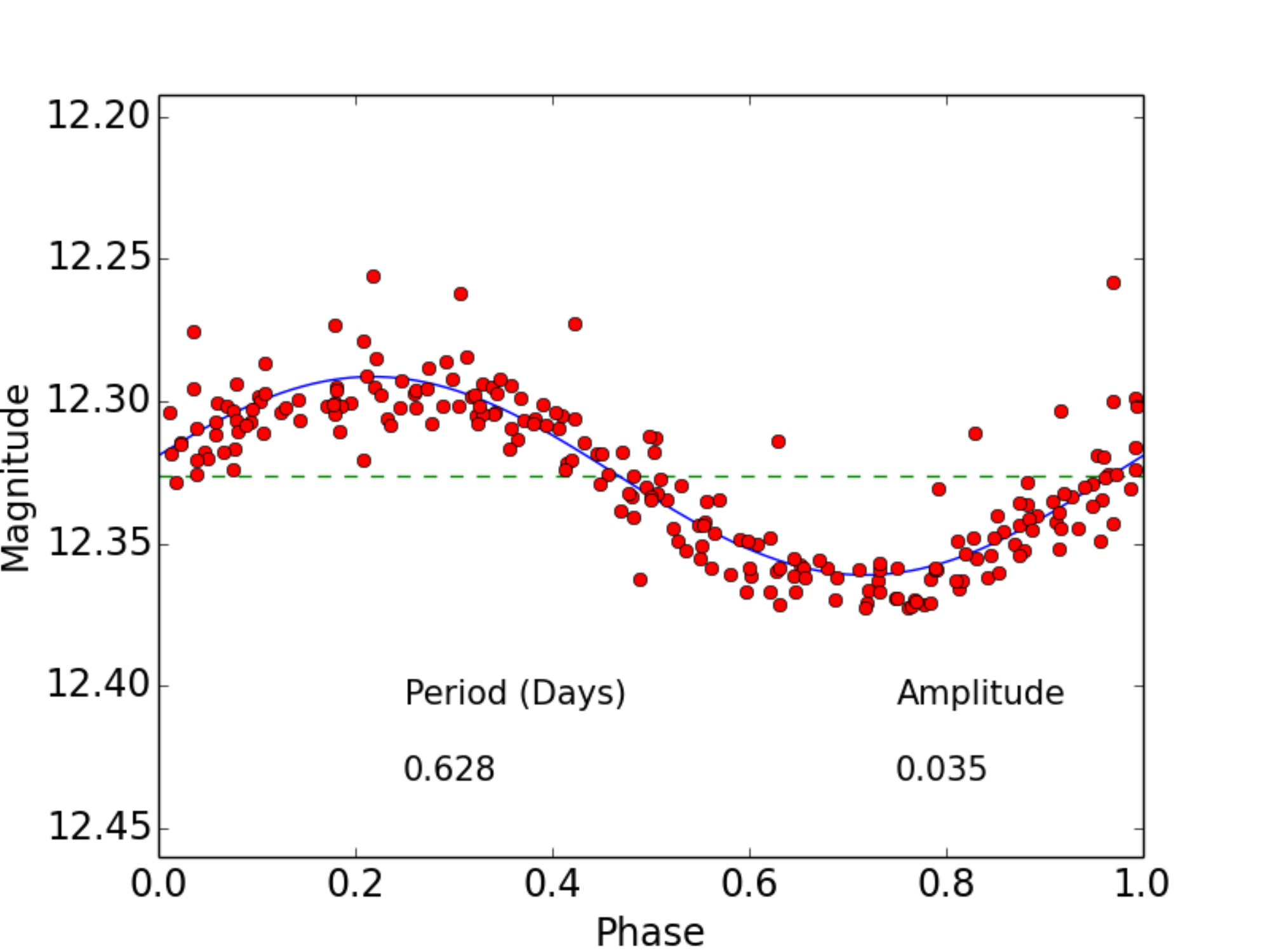}
\caption{Periodograms (top figures) and phase-folded light curves
  (bottom figures) for three stars, illustrating a range of periods
  and amplitudes. The dashed red line represents the power at the 0.01
  FAP and the adopted frequencies for each star are marked with a
  black arrow. In the light curves, best-fit sinusoid is plotted in
  blue and the green dashed line represents the SWASP mean magnitude.
  {\it Left}: 2MASS J15082502-3337554, {\it Center}: 2MASS
  J13191370-4506326, {\it Right}: 2MASS J11132622-4523427.}
\label{result_examples}
\end{figure*}

In short, the FAP for a given star was estimated by randomly shuffling
its light curve’s photometric values relative to its time values
(after binning), creating a new, randomized light curve. Next, a
periodogram was created for this randomized light curve and its
strongest period saved. This was repeated $10^4$ times, and the
resulting strongest periods were plotted as a histogram. Finally, the
strongest period from the \textit{real} light curve was compared to
this histogram. The period was considered to be a real detection only
if less than 10\% of the histogram’s periods had stronger peaks. In
this case we say the the FAP $<$ 0.01. Obvious aliases (periods
occurring at integer multiples of the strongest period) were vetted by
eye. {Figure \ref{result_examples}} shows three example light curves
with their corresponding periodograms and FAP levels.

%%% FIGURE 5: Periodogram Plots The

Uncertainties for the periods were measured only for the stars
presented in {Table \ref{Tab:Results}} and {Table
  \ref{Tab:Other_Stars}}. The maximum peak in the periodogram for each
season was fit with a Gaussian using a least-squares routine. The
means and variances were averaged for each season and reported in
{Table \ref{Tab:Results}} and {Table \ref{Tab:Other_Stars}}. The light
curves for each star was then phase-folded over its seasonal period
and plotted.
Using the {\it scipy} least-squares fitting tool, a sine curve is fit
to each phase-folded light curve ({Figure \ref{result_examples}}).
%%% SEC 3.2

{Figure \ref{result_examples}} contains example periodograms and light
curves for three example stars in our sample.
The criteria for a detected rotation period are (1) a periodogram peak
must be above the 0.01 FAP plot threshold, (2) that said peak is
consistent through $\textit{at least two}$ seasons of SuperWASP data,
and (3) a consistent light curve amplitude (to within a few 0.1
magnitudes) through at least two seasons.
Further, the stars reported must be confirmed members of Sco-Cen (see
{\S \ref{Sec:Time_Photom}}).
In order to completely profile each star and compare them onto angular
momentum evolution models (see { Figures \ref{P-M-Gallet} \&
  \ref{P-A-Gallet}}), the star must have a published spectral type and
be a member of one of the three main subgroups.

\subsection{Intrinsic Color, Temperature, and Mass Modeling}
\label{Sec:Intrinsic} 

\begin{figure*}[h]
\centering
\includegraphics[width=1\textwidth]{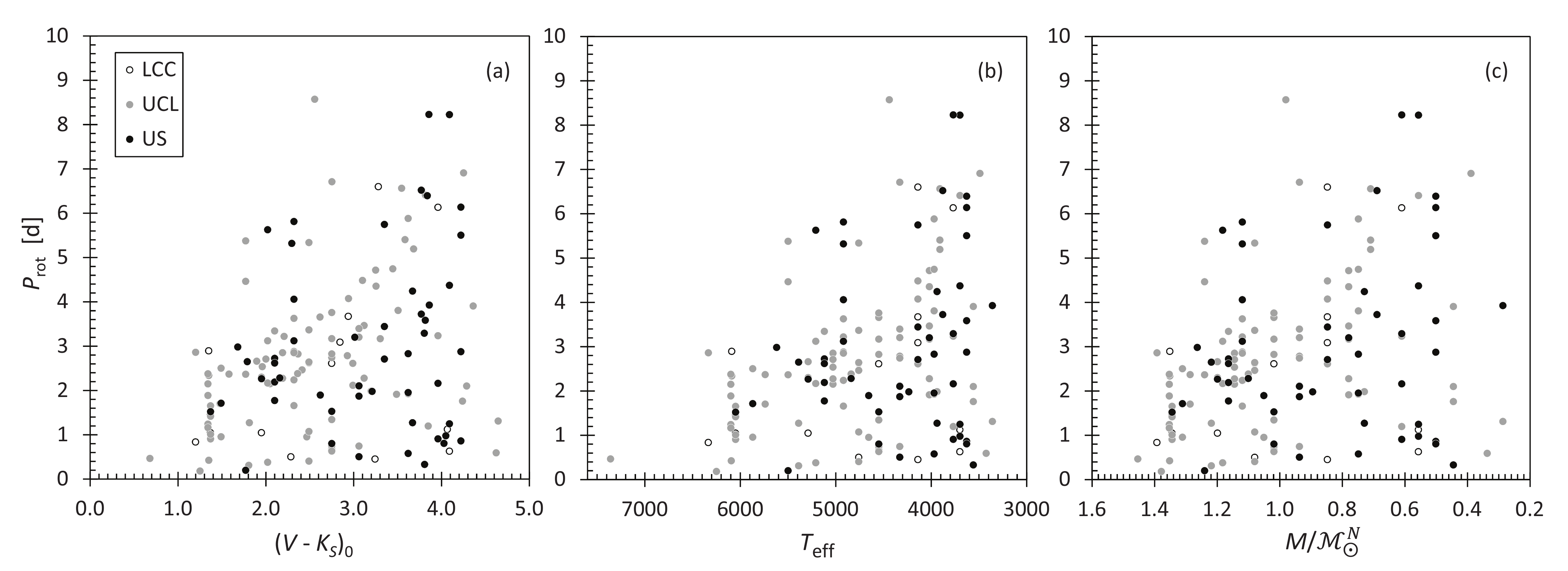}
\caption{Period-color, -temperature, and -mass diagrams. The vertical
  axes plot the season-averaged rotation period determined from the
  SuperWASP periodogram analysis in units of days, \prot. The
  horizontal axes plot the extinction-corrected observed color \vks\ in
  magnitudes (Panel a), the modeled effective temperature \teff\ in
  units of K (Panel b), and the modeled mass in units of
  solar masses, $M$/\Msun\, (Panel c). Panel (a) contains the complete
  final sample of 162 stars in the classic Sco-Cen subgroups of US
  (filled black circles), UCL (filled gray circles), and LCC (open
  circles), whereas panels (b) and (c) contain only the 157 of these
  stars with measured spectral-types --- since the temperature and mass
  models depend on the spectral type.\label{Fig:Period-Color}}
\end{figure*}

To investigate the angular momentum evolution of Sco-Cen stars, it is
constructive to plot the rotation periods determined from the
periodogram analysis as a function of intrinsic stellar properties.
In particular, the measured spectral types for 157 stars in our sample
enabled us to estimate intrinsic colors (\vks), effective temperatures
(\teff), and masses ($M$).
The intrinsic \vks\, and \teff\, were determined from the empirical
relations and BT-Settl grid models, respectively, from Table 6 of
\citet{Pecaut13}.
Based on these \teff\, values, we computed $M$ using the theoretical
isochrones of \cite[][hereafter BHAC15]{Baraffe15} (adopting the mean
age of each star's subgroup):

\begin{equation}
\begin{footnotesize}
\begin{array}{ll}
\log(M/\mathcal{M}^{\rm N}_{\odot}) = & - \, (3.5075918704 \times 10^4)\\
	& + \, (3.7421144677 \times 10^4)\cdot\log(T_{\rm eff})\\
        & - \, (1.4972186556 \times 10^4)\cdot\log(T_{\rm eff})^2\\
        & + \, (2.6625089056 \times 10^3)\cdot\log(T_{\rm eff})^3\\
        & - \, (1.7755725859 \times 10^2)\cdot\log(T_{\rm eff})^4,
\end{array}
\end{footnotesize}
\end{equation}

\noindent this is valid over the range 3056 K $<$ \teff\, $<$ 6422 K
and 0.1 $<$ $M$/\Msun\, $<$ 1.4, and has a calibration uncertainty of
approximately $\sigma$(\logm) $\simeq$ 0.007 dex.
% (Only four stars in our sample fall outside of this temperature
% range, having values of \teff\, = 6590, 6625, 6990, and 7360 K).
Table \ref{Tab:Results} lists the intrinsic \vk, \teff, and masses
determined for the stars in our sample, and period-color,
period-temperature, and period-mass diagrams are plotted in Figure
\ref{Fig:Period-Color}.

%%% FIGURE 6 Period-Color Plot

%%% SEC 3.2, PAR 
Interstellar reddening was estimated using $B-V$, $V-J$, $V-H$, and
\vk.  The observed colors were computed from $BV$ photometry collected
from various sources (e.g., APASS DR9, Hipparcos, Tycho-2, or other
ground-based photometry) and near-infrared photometry from 2MASS
\citep{Cutri03}.
The photometry adopted for estimating 
extinctions is listed in {Table \ref{Tab:Pre-AnalysisShort}}.
The observed photometry was then compared to intrinsic color sequences
for pre-MS stars as a function of spectral type from \citet{Pecaut13}
to estimate color excesses $E(B-V)$, $E(V-J)$, $E(V-H)$, and
$E(V-K_S)$.
Extinctions for individual objects were computed using these four
color excesses, as described in \citet{Pecaut16}.  The computed
extinctions for each star are listed in {Table \ref{Tab:Results}}.

For stars without a measured spectral type, the color excess and
extinction coefficient were taken as the average of the other stars in
the same subgroup: E(V-K$_S$)$_{avg}$ = 0.332 for US and 0.173 for
each of UCL and LCC; (A$_V$)$_{avg}$ = 0.372 for US and 0.194 for each
of UCL and LCC.
These mean color excesses and extinctions were used to correct
the observed $V$ and \vk\, values in the color-magnitude diagram of
Figure \ref{Fig:Color-Mag} and the observed \vk\, colors in the
period-color plot of Figure \ref{Fig:Period-Color} Panel (a).

\section{Discussion} %%%% SEC 4

\subsection{Results}
\label{results}

{Table \ref{Tab:Results}} summarizes the relevant stellar parameters
for our sample stars, including SWASP -- 2MASS cross-identifications,
Sco-Cen sub-group assignments, season-averaged rotation periods,
spectral types, colors, extinctions, effective temperatures, and
estimated masses.
A search of the literature found only one variability study of stars
across Sco-Cen with which to compare \citep{David14}.
We find that the only star with a reported period in both our survey
and \citet{David14} was HD 141277 (2MASS J15494499-3925089), with both
surveys reporting the same rotation period of $\sim$ 2.23 days.
We queried the AAVSO International Variable Star Index (VSX) catalog
via Vizier using the 2MASS positions of our stars, with a
5\arcsec\ search radius; the search uncovered the nearest spatial
matches (most $<$ 1\arcsec\ away) for 94 of our stars \citep{VSX}.
Of these 94 stars, 76 have periods similar to ours.
There are 96 new rotation period measurements, including a second,
shorter activity period measured for the Li-rich K giant CD-43 6891.
\subsection{Rotational Evolution}
\label{rot_ev}

%%% FIGURE 7: MASS VS. PERIOD ACROSS WIDE MASS RANGE (LIMITS: 0.2-1.6 MSUN)

\begin{figure*}[h]
\centering
\includegraphics[width=0.75\textwidth]{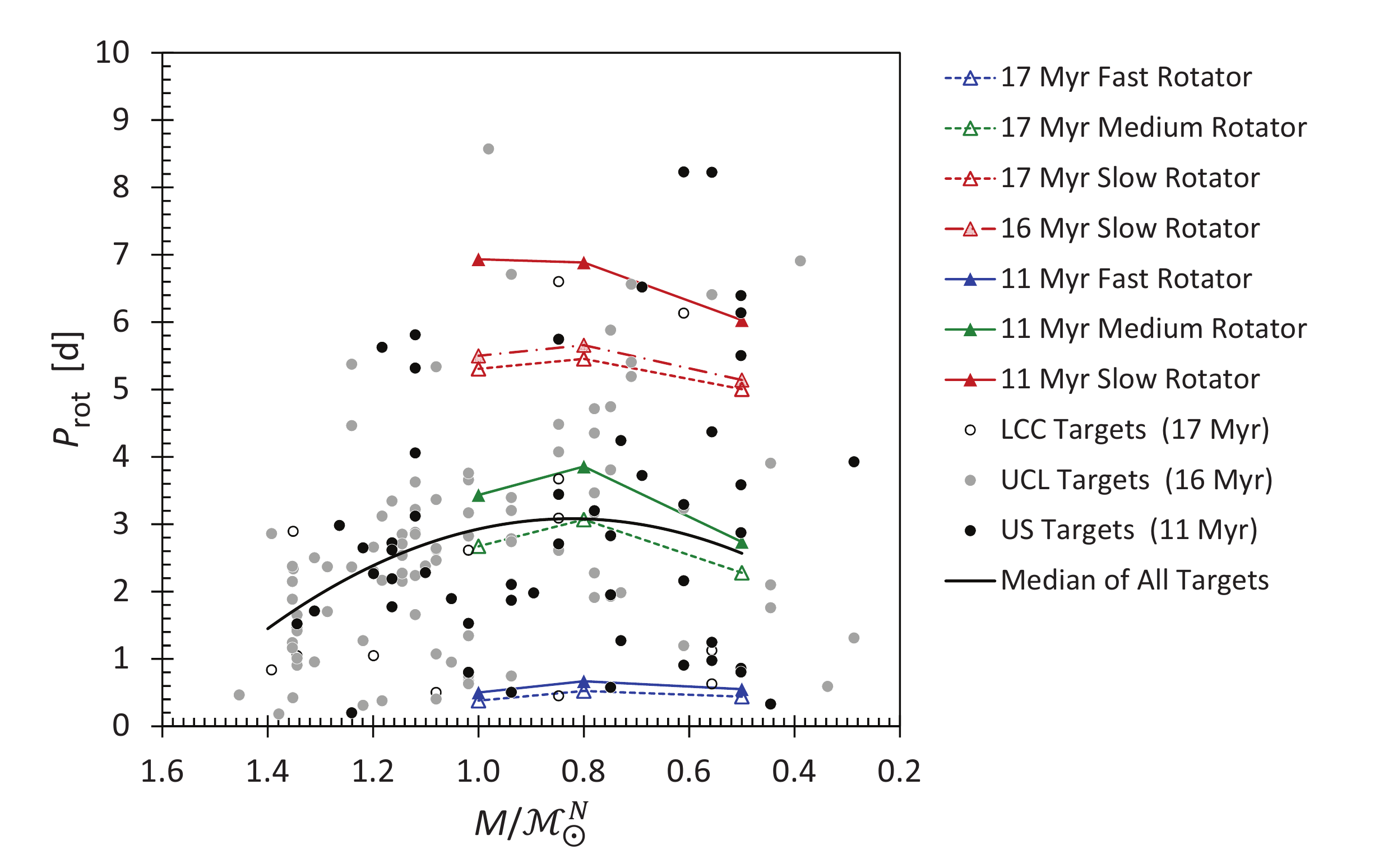}
\caption{The period-mass diagram of Figure \ref{Fig:Period-Color}
  panel (c) overlaid with the angular momentum evolutionary model
  tracks of \citet{Gallet15}. As in \citet{Gallet15}, tracks are shown
  for each of ``slow,'' ``medium,'' and ``fast,'' rotators
  corresponding to the 25$^{\textrm{\scriptsize{th}}}$,
  50$^{\textrm{\scriptsize{th}}}$, and 90$^{\textrm{\scriptsize{th}}}$
  percentiles in rotational period of the stellar envelope. The
  modeled points (triangles) have been averaged by mass in bins of 0.4
  - 0.6, 0.7 - 0.9, and 0.9 - 1.1 \Msun\, and are plotted at the mass
  bin centers; the connecting lines serve only to interpolate between
  these modeled points and do not themselves represent model
  data. Here only model tracks corresponding to the median ages of US,
  UCL, and LCC of 11, 16, and 17 Myr, respectively, are
  reproduced. Model tracks for the 16 Myr medium and fast rotators
  have been omitted since they are nearly identical to (i.e. overlap)
  those of the 17 Myr medium and fast rotators, respectively. For
  general comparison to the model tracks, a second-order polynomial
  (black solid line) was fit to the median periods of the 157 stellar
  data points (circles) in same three mass bins as the models plus two
  additional bins of 1.1 - 1.3 and 1.3 - 1.5 \Msun.}
\label{P-M-Gallet}
\end{figure*}

%%% FIGURE 8: AGE VS. PERIOD FOR DIFFERENT MASS BINS
\begin{figure*}[h]
  \centering
  \includegraphics[width=1\textwidth]{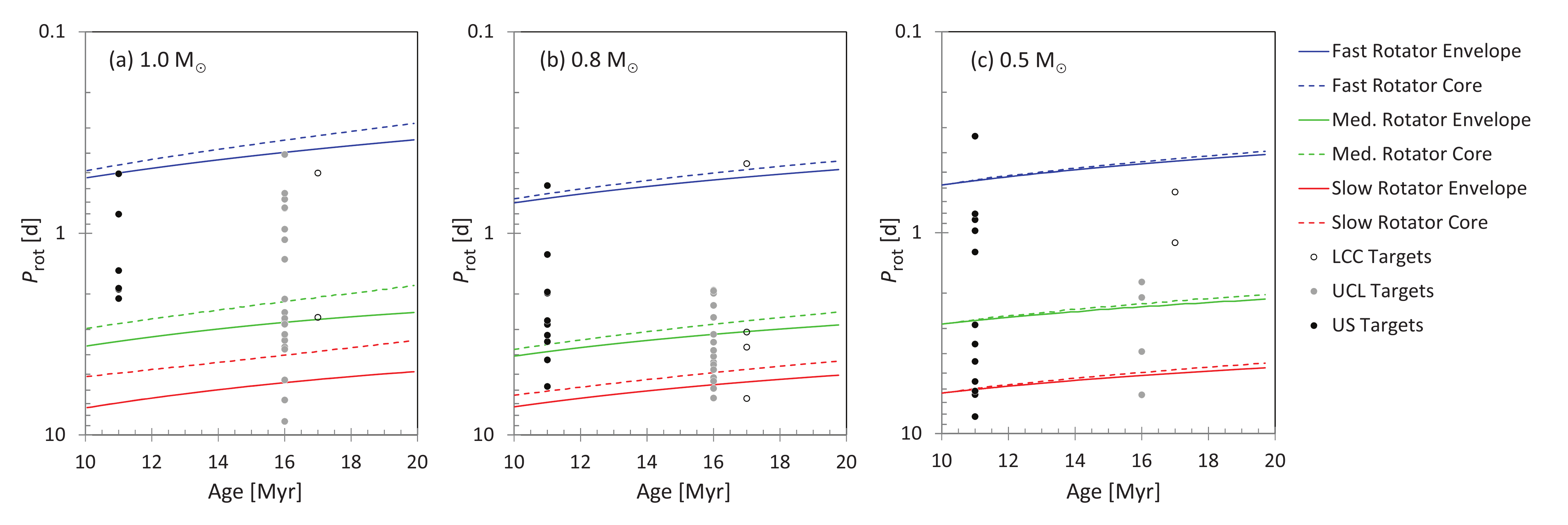}
  \caption{A reproduction of the period-age diagrams of
    \citet{Gallet15}, Figure 5, with Sco-Cen stars from the present
    study (circles) overplotted. As in \citet{Gallet15}, angular
    momentum evolutionary model tracks are shown for each of ``slow,''
    ``medium,'' and ``fast,'' rotators corresponding to the
    25$^{\textrm{\scriptsize{th}}}$, 50$^{\textrm{\scriptsize{th}}}$,
    and 90$^{\textrm{\scriptsize{th}}}$ percentiles in rotational
    period for each of the stellar core and envelope. The modeled data
    are also averaged by mass in bins of 0.4 - 0.6 \Msun\, (panel a),
    0.7 - 0.9 \Msun\, (panel b), and 0.9 - 1.1 \Msun\, (panel c). Only
    82 of the final 162 star sample are shown since five do not have
    mass determinations and the rest have masses falling outside of
    these three mass ranges.}
\label{P-A-Gallet}              
\end{figure*}

For each star, we adopt the mean age of its subgroup and use its
spectral type (and corresponding \teff) to estimate a mass using
isochrones (see {\S \ref{Sec:Intrinsic}}).
{Figure \ref{P-M-Gallet}} is a mass-period diagram for Sco-Cen
with rotational evolution tracks from \citet{Gallet15} overplotted for
several rotation rates at the ages of each major subgroup.
In addition, a quadratic is plotted in an attempt visualize the
rotational evolution trend of the $\sim$ 11-17 Myr sample of pre-MS
stars.
The season-averaged rotation periods versus the age of each subgroup
are plotted in {Figure \ref{P-A-Gallet}}.
The \citet{Gallet15} rotational evolution tracks are overplotted for
different stellar masses and initial conditions (fast, medium, slow
rotators); initial conditions were chosen to best match rotation
period distributions of older and younger clusters.
Our data show that the general trend in period and mass predicted
by the \citet{Gallet15} models continue to do a reasonable job
matching the Sco-Cen stars.
The envelope of the fastest rotators in Sco-Cen between $\sim$ 0.5 --
1 \Msun\, follows the fast rotator trend predicted by
\citet{Gallet15}.
We show a few stars with periods of $\sim$ 8 days that are rotating
slightly slower than the envelope of the slowest rotators predicted by
the \citet{Gallet15} models.
All of our $\sim$ 1.25 -- 1.5 \Msun\, stars are rotating very fast
(0.2 -- 3 days) and just below $\sim$ 1.25 \Msun\, stars are found
with much slower rotation periods of $\sim$ 5 -- 6 days.

These results suggests that while the stars that will eventually be
early F dwarfs on the main sequence (F1V-F6V; 1.25 -- 1.5 \Msun) have
rotation periods at $\sim$ 11 -- 17 Myr that on average are only
slightly less than their G-type brethren ($\sim$ 2 days vs. $\sim$ 3
days, on average), the F dwarf population appears to arrive on the
main sequence with a population near breakup velocity (P $\simeq$ 0.2
day; e.g. 1SWASP J154610.69-384630.2, 1SWASP J144619.03-354146.5), but
lacking a population of slow rotators (P $>$ 3 days). The early F-type
stars remain fast-rotating throughout their main sequence lifetimes,
experiencing only weak magnetic breaking due to their weaker magnetic
dynamos \citep[e.g.][]{Slettebak55,Kraft67}.

\subsection{Search for Eclipsing Systems}

In addition to our rotation evolution study, this analysis included a
search for any signatures of circumsecondary disks similar to 1SWASP
J140747.93-394542.6 \citep[``J1407'' = V1400 Cen;][]{Mamajek12,Scott14}.
The asymmetric light curves of the stars $\epsilon$ Aurigae
\citep{Guinan02}, EE Cep \citep{Mikolajewski99}, and J1407 are
signatures of disk or ring systems occulting their host stars.
These asymmetric eclipses are characterized by their long duration and
magnitude-scale depths \citep[e.g. EE Cep with a $\sim$ 30--90 day
  eclipse and depth of $>$ 2 mag;][]{Mikolajewski99}.
An effective large scale search for these eclipses \citep[as proposed
  by][]{Mamajek12} would require a long term (10 year), high cadence
time-series photometric survey of $10^{4}$ post-accretion \prems\,
stars --- yielding only a few candidates.

All of the light curves from this analysis were scanned by eye for
obvious eclipses like J1407 (see {Figure \ref{J1407_LC}}).
To see an eclipse occur, it would have to be restricted to one of these
100 day windows \citep[J1407 has a $\sim$ 58 day eclipse
  duration;][]{Mamajek12}.
Additionally, the rotation period amplitudes vary on the 0.01
magnitude level (see {Table \ref{Tab:Photometry}}).
Thus, any amplitude variations on the 0.1 magnitude level would be
evident in our periodogram search, making shorter term $<$ 10 day
eclipses evident.
Therefore, no eclipses with durations of $<$ 100 days and depths $>$
0.1 magnitudes were detected in this sample.

This survey also uncovered five candidates for pre-MS eclipsing
binaries.
However, after further review of their astrometric data, we reject
Sco-Cen membership for four of them.
We argue that one of them (V2394 Oph) appears to be a poorly
characterized, heavily reddened (A$_V$ $\simeq$ 5 mag) massive
eclipsing binary in the LDN 1689 dark cloud of the Ophiuchus
star-forming region of Sco-Cen.
Further discussion on these eclipsing binaries can be found in the
Appendix.

\subsection{Conclusion}

This survey searched the SuperWASP archive for Sco-Cen members with
measurable rotation periods and significant eclipsing events. A total
of 189 reliable rotation periods were extracted -- 162 for stars in
the classic Sco-Cen subgroups of US, UCL, and LCC, and 27 for stars in
younger star-forming regions within the Sco-Cen complex. Of these, 157
of the classic subgroup members have previously reported spectral
types. These spectral types were used to estimate masses to compare
our data against current angular momentum evolution models from
\citet{Gallet15}, and the rotation periods appear to be in reasonable
agreement with the range of periods predicted by the models. No new
eclipsing circumsecondary disks were detected beyond the previously
known V1400 Cen (J1407) system.  Five eclipsing binary systems were
identified, but only one appears to be a strong candidate for
membership in Sco-Cen (V2394 Oph in the LDN 1689 dark cloud in
Ophiuchus).  The remaining four eclipsing binaries all appear to be
interlopers.

\section*{Acknowledgements}

SNM was supported by NSF award PHY-1156339 for the University of
Rochester REU program.
EEM acknowledges support from NSF award AST-1313029.
SNM and EEM also acknowledge support from the NASA NExSS program.
This work used the Vizier and SIMBAD services.
Part of this research was carried out at the Jet Propulsion
Laboratory, California Institute of Technology, under a contract with
the National Aeronautics and Space Administration.
We also acknowledge and thank Florian Gallet for providing the angular
momentum evolution tracks used to generate {Figures \ref{P-M-Gallet}
  \& \ref{P-A-Gallet}}.
This paper makes use of data from the first public release of the WASP
data \citep{Butters10} as provided by the WASP consortium and services
at the NASA Exoplanet Archive, which is operated by the California
Institute of Technology, under contract with the National Aeronautics
and Space Administration under the Exoplanet Exploration Program.
This document has been approved for unlimited release (CL\#17-2577).

\bibliographystyle{aasjournal} % NOT APJ.BST! 
%\bibliography{mamajek}{}
\bibliography{mamajek}
\appendix

\section{Periods for Other Stars in Sco-Cen Complex}

SuperWASP time series photometry was found for 27 other stars. These
stars are associated with neighboring subgroups (some of which are
active star-forming regions, e.g. Lup, Oph, CrA), i.e. their positions
and/or kinematics are inconsistent with membership within the three
classic older subgroups (LCC, UCL, US). Two are associated
  with the TW Hya group. We also found a Li-rich K giant with two
  measured periods, which is elaborated on later in the appendix.
Their estimated rotation periods (calculated following the analysis in
{\S 3}), along with their 2MASS alias and spectral type are presented
in {Table \ref{Tab:Other_Stars}}.

\section{A Massive Eclipsing Binary in Ophiuchus: V2394 Oph}

{\it 1SWASP J163140.67-242516.2} (V2394 Oph, TYC 6799-309-1, CoD-24
12698) is a 0.59 day eclipsing binary whose components are probably
either in contact or close. The star is situated in the LDN 1689 cloud
\citep{Nutter06}, and has been previously selected as a proper motion
member of either Upper Sco \citep{Hoogerwerf00} and/or Oph
\citep{Makarov07}.  A very strong period of 0.295 day is clearly
detected in all three seasons of SuperWASP data.  Phase-folded time
series photometry at 0.295 day suggests that the secondary eclipse
depth is $\sim$ 0.2 mag and the primary eclipse depth is $\sim$ 0.4
mag. The long-term out-of-eclipse brightness seems to be varying at
the $\sim$ 0.1 mag level over the three years.  \citet{Grankin96}
report the eclipsing binary to have photometry at maximum of $V$ =
10.01, $U-B$ = 0.40, $B-V$ = 0.95.  \citet{Barnard10} and
\citet{Struve62} both commented on the concentration of nebulosity
centered on the star (called CD -24$^{\circ}$~12684 in these
publications), and \citet{Struve62} comments on a very red {\it
  reflection nebula} surrounding the star and reports that a spectrum
taken by George Herbig in 1949 revealed the star to be A0 or
A1. \citet{Vrba76} reported the star (VSS II-50) as spectral type B9,
and estimated the star to have extinction $A_V$ $\simeq$ 3.29
(E($B-V$) = 1.06).  The \citet{Grankin96} colors at maximum light are
a good match to a \teff\, $\simeq$ 9700\,K dwarf with E($B-V$) = 1.47
($A_V$ = 4.87 mag), which would be consistent with $\sim$ A0 type.

Is this star associated with Sco-Cen? The Gaia DR1 TGAS astrometry for
star lists proper motion $\mu_{\alpha}$, $\mu_{\delta}$ = -6.672,
-24.594 ($\pm$1.599, $\pm$1.803) mas\,yr$^{-1}$ and parallax $\varpi$
= 7.23\,$\pm$\,0.58 mas (consistent with distance 138\,$\pm$\,11
pc). The proper motion is similar to the mean proper motion of the
YSOs in the Oph embedded cluster: $\mu_{\alpha}$, $\mu_{\delta}$ =
-10, -27 ($\pm$2, $\pm$2) mas\,yr$^{-1}$ \citep{Mamajek08Oph}.  The mean
distance to the Oph cluster has been estimated recently to be
131\,$\pm$\,3 pc \citep{Mamajek08}, 119\,$\pm$\,6 pc
\citep{Lombardi08}, 120.0$^{+4.5}_{-4.2}$ pc \citep{Loinard08}, but a
recent VLBA trigonometric parallax survey of 16 systems by
\citet{OrtizLeon16} refined the mean distance to 137.3\,$\pm$\,1.2 pc.
Hence, both the Gaia DR1 TGAS proper motion and parallax are
statistically consistent with the rest of the YSO population in the
Oph cloud. If the star is associated with Oph, its radial velocity is
predicted to be -6.5\,km\,s$^{-1}$. The star is in very close
proximity with 4 other lower-mass YSOs within 3' ($\sim$ 0.12 pc
projected radius; DoAr 43, 44, 46, and 2MASS J16313124-2426281), which
may consist of an unstable dynamical trapezium.

Hence, V2394 Oph appears to be not only comoving with Upper Sco and
Oph, in the immediate vicinity of other YSOs, associated with
nebulosity, and a Gaia DR1 parallax consistent with being co-distant
with the Oph clouds \citep[see clumping of parallaxes for stars
  illuminating reflection nebulae in ][]{Mamajek08}.  The photometry
is consistent with an unresolved, unextincted $V_o$ magnitude of
5.14. Using the TGAS parallax, we estimate an unresolved absolute
magnitude of $M_V$ = -0.6, which places it about 2 mag above the
zero-age main sequence \citep{Aller82}. Through comparison with the
\citet{Siess00} isochrones, and considering the unknown mass and radii
ratios of the components, we propose that the V2364 Oph system is a
$\sim$ 1-2 Myr-old contact or near-contact eclipsing binary where the
primary is a $\sim$ 3-5 $M_{\odot}$ star.  \citet{Erickson11} only
identify four members of the Oph clouds whose spectral types are A0 or
earlier (HD 147889, SR 3, Oph S1, WLY 2-48), and now V2394 Oph appears
to be the most massive member specifically of the LDN 1689 dark
cloud.

\section{Other Eclipsing Binaries}

{\it 1SWASP J140807.34-393548.8 (TYC 7807-358-1, ASAS J140807-3935.9,
  CD-39 8717)} appears to be a 7.815 day eclipsing binary with primary
eclipse depth of $\sim$ 0.7 mag and secondary eclipse depth of $\sim$
0.6 mag. The strongest peaks have periods of 7.83 days (the season 1
phase-folded light curve shows two minima), 3.91 days (the season 2
single minimum), and 3.895 days (the season 3 single minimum with
large scatter). This star was selected as a candidate UCL member by
virtue of its proper motion by \citet{Hoogerwerf00}. The UCAC4 proper
motion is suggestive of LCC membership, however the kinematic parallax
($\varpi$ = 9.30\,$\pm$\,0.62 mas) we calculate using the UCAC4 proper
motion and LCC space motion from \citet{Chen11} differs from the Gaia
DR1 parallax ($\varpi$ = 2.98\,$\pm$\,0.75 mas) by 6.5$\sigma$. We
conclude that this is a background interloper unrelated to Sco-Cen.\\

{\it 1SWASP J151126.76-361457.2 (HD 134518)} is a 1.154 day eclipsing
binary with a primary eclipse depth of $\sim$ 0.15 mag and secondary
eclipse depth of $\sim$ 0.05 mag. Very strong peaks at exactly 0.577
days are detected in each of the three seasons. This star was
mentioned as a UCL candidate by
\citet{deGeus89}. \citet{Perevozkina04} classifies the system as
A7+[K5], and \citet{Houk82} classifies the blended spectrum as A8V.
The revised Hipparcos parallax \citep[$\varpi$\,=\,7.72\,$\pm$\,1.81
  mas][]{vanLeeuwen07} and light estimated reddening (E($B$-$V$)
$\simeq$ 0.09 are both similar to other UCL members (mean $\varpi$
$\simeq$ 7.1 mas), however the revised Hipparcos proper motion is off
of the mean UCL motion \citep{Chen11} by $\sim$ 5\,km\,s$^{-1}$ - much
larger than the 1D velocity dispersion of the group ($\sim$ 1.3
km\,s$^{-1}$). The Hipparcos parallax translates the primary's HRD
position to on the main sequence and below the trend for other Sco-Cen
members. We consider HD 134518's membership to UCL unlikely, but
further follow up is warranted.  If HD 134518 belongs to UCL, its
systemic radial velocity should be 3 km\,s$^{-1}$.\\

{\it 1SWASP J153554.13-335623.6 (TYC 7322-822-1)} is a 1.06 day
eclipsing binary showing a primary eclipse depth of $\sim$ 0.3 mag and
secondary eclipse depth of $\sim$ 0.2 mag. The system showed very
clean 0.53 day peaks in all three seasons. \citet{Hoogerwerf00}
selected the star as a candidate UCL member based on its proper
motion. The star's Gaia DR1 parallax ($\varpi$ = 3.19\,$\pm$\,0.78
mas) differs from the kinematic parallax we calculate \citep[$\varpi$
  = 8.52\,$\pm$\,0.70 mas; adopting its UCAC4 proper motion and
  assuming UCL space velocity from][]{Chen11} by 5.1$\sigma$. We
conclude that this is a background interloper.\\

{\it 1SWASP J154856.93-363920.2 (TYC 7340-720-1)} is a 0.393 day
contact binary with primary eclipse depth of $\sim$ 0.10 mag and
secondary eclipse depth of $\sim$ 0.08 mag.  \citet{Hoogerwerf00}
selected this star as a candidate UCL member based on its proper
motion. The Gaia DR1 parallax ($\varpi$ = 6.51\,$\pm$\,0.32 mas) is
similar to other UCL members, however the Gaia TGAS proper motion is
off of the predicted UCL motion by 8\,$\pm$\,1 mas\,yr$^{-1}$
(7\,$\pm$\,1 km\,s$^{-1}$). We conclude that the star is an
interloper.

\section{A Li-Rich Red Giant Interloper}

{\it 1SWASP J111434.43-441824.1 (2MASS~J11143442-4418240, CD-43 6891)}
gave conflicting signals regarding its potential Sco-Cen membership.
The star was found by \citet{Pecaut16} to be a Li-rich (EW(Li I
$\lambda$6707) = 670 m\AA) X-ray-emitting K2IV star, however its
inferred isochronal age (1 Myr) appeared to be extraordinarily young.
The star has a large photometric amplitude ($\sim$ 0.2 mag) and long
period (37 day), for which \citet{Richards12} classified its light
curve as that of a small-amplitude red giant type B.  Both
\citet{Rybka07} and \citet{Gontcharov08} flag the star as being a
likely clump red giant.  The new Gaia DR1 parallax \citep[$\varpi$ =
  2.18\,$\pm$\,0.25 mas;][]{Gaia2016} is clearly at odds with the
predicted kinematic parallax calculated by \citet[][; $\varpi$ =
  6.56\,$\pm$\,0.60 mas; which assumed Sco-Cen membership]{Pecaut16}.
Ignoring the effects of extinction, this translates to an absolute
magnitude of M$_V$ $\simeq$ 1.5, which puts it squarely among other K2
giants.  We conclude that the star is a rare Li-rich giant, and an
interloper.

%%% FIGURE 9: Phase-folded LC's of EB's
\begin{figure*}[h]
\centering
\includegraphics[width=.4\textwidth]{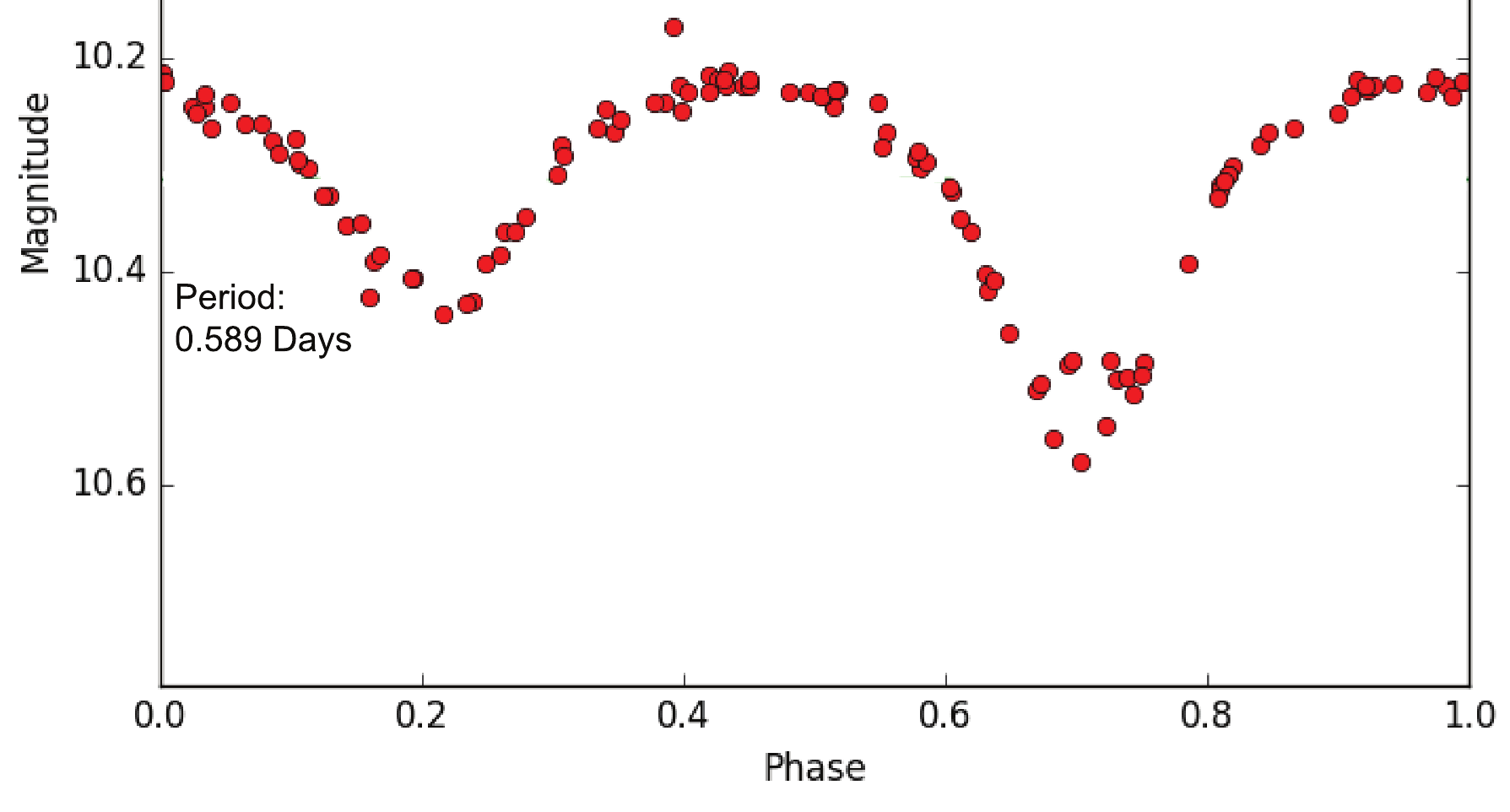}
\includegraphics[width=.4\textwidth]{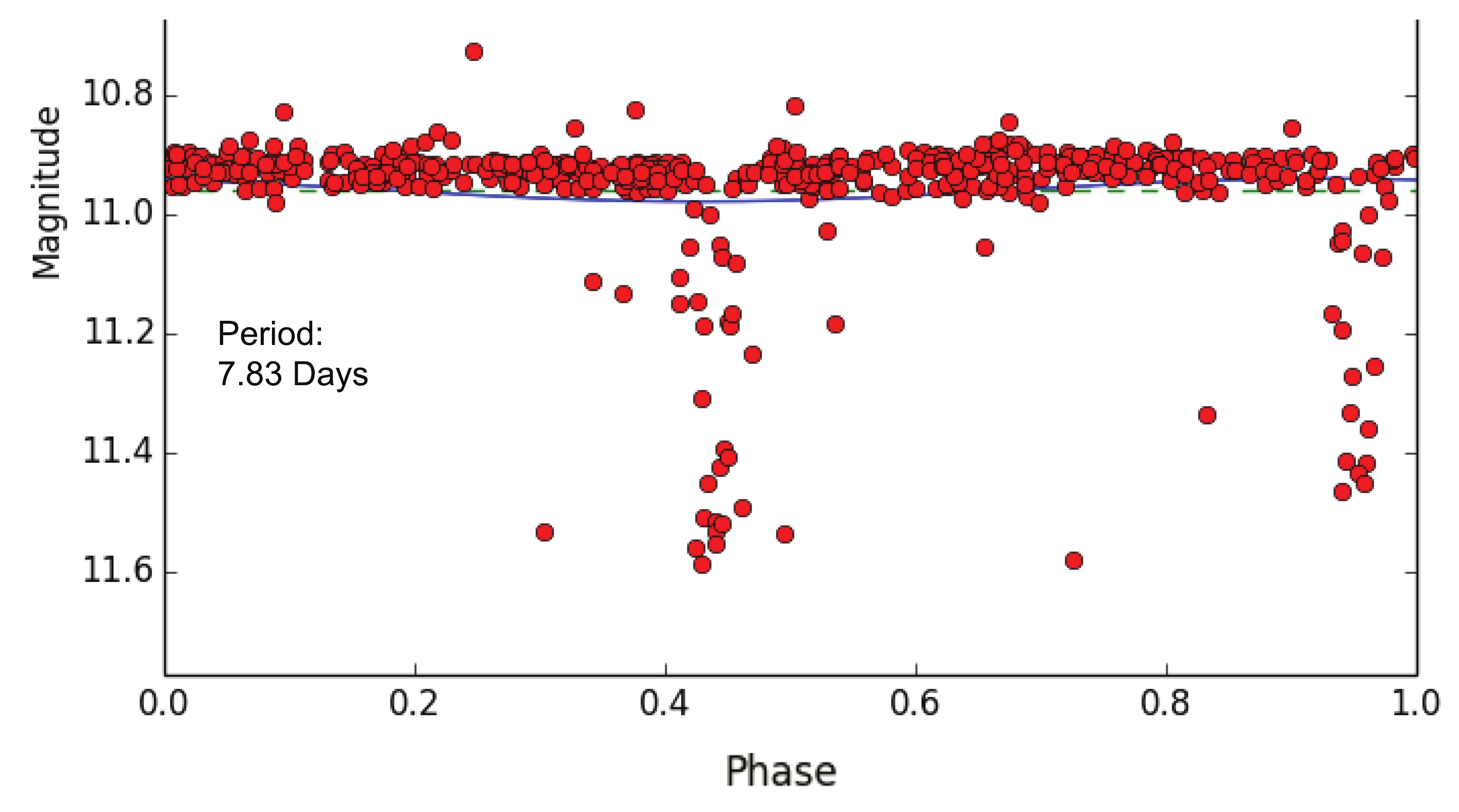}\\
\includegraphics[width=.4\textwidth]{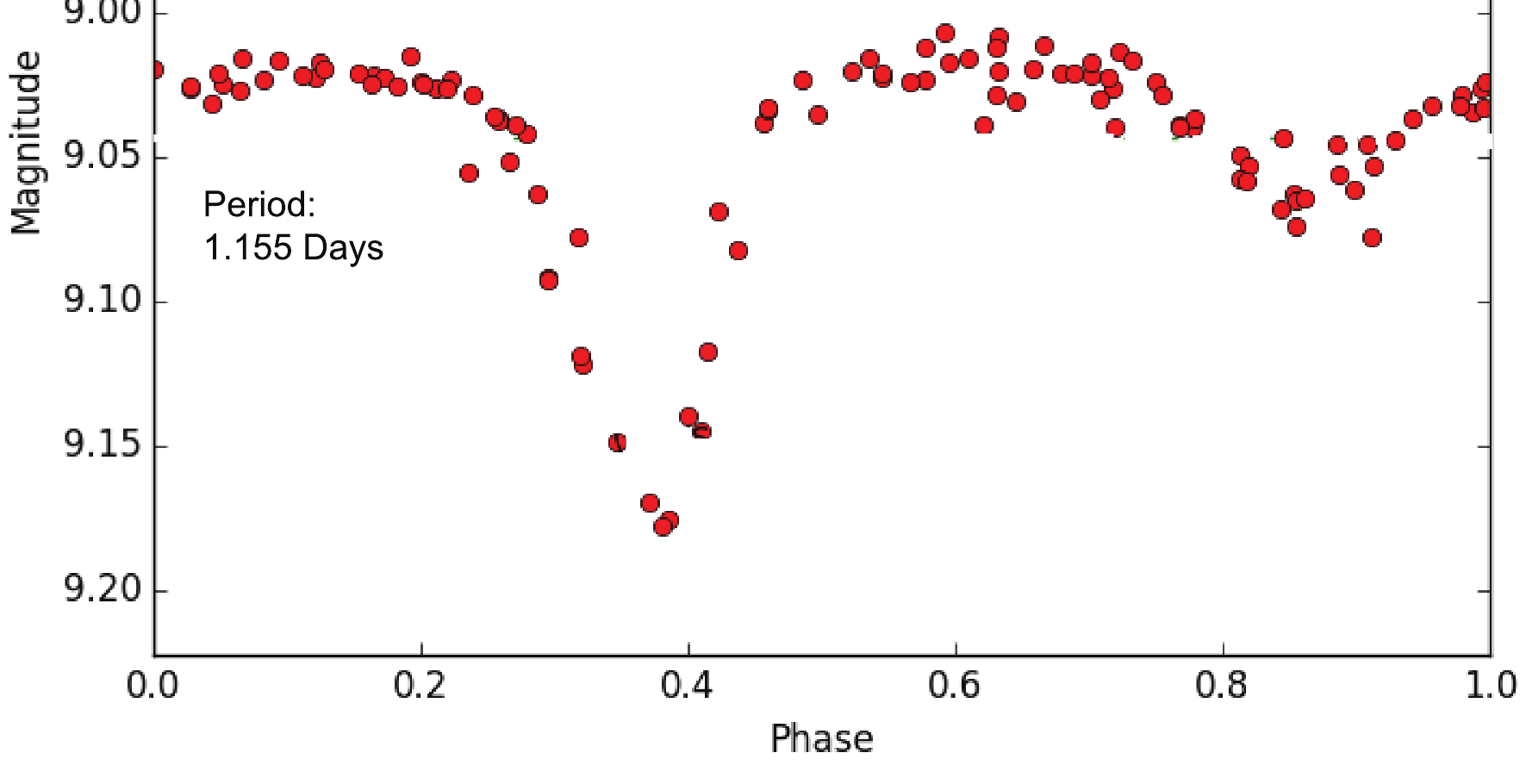}
\includegraphics[width=.4\textwidth]{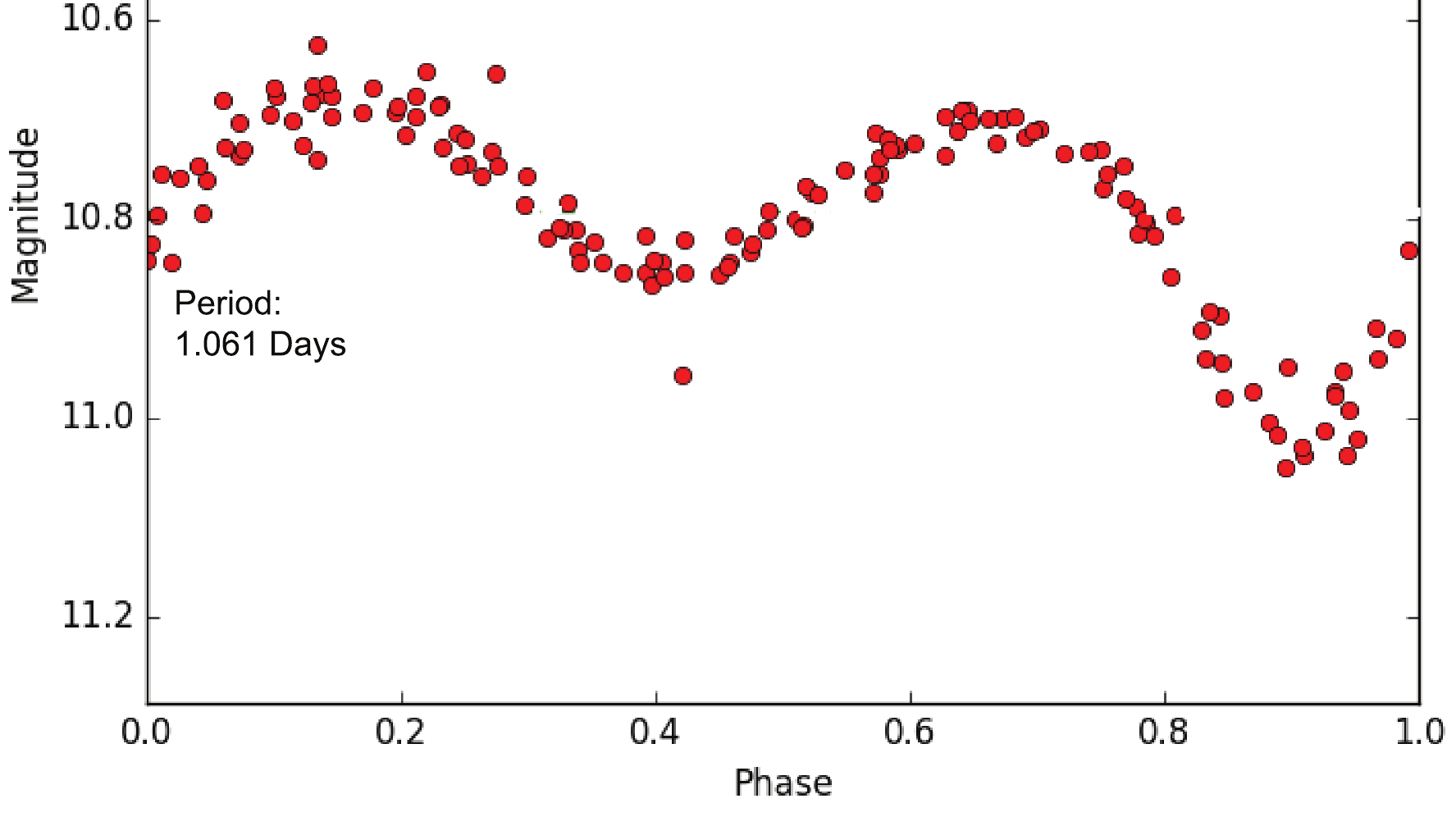}
\includegraphics[width=.4\textwidth]{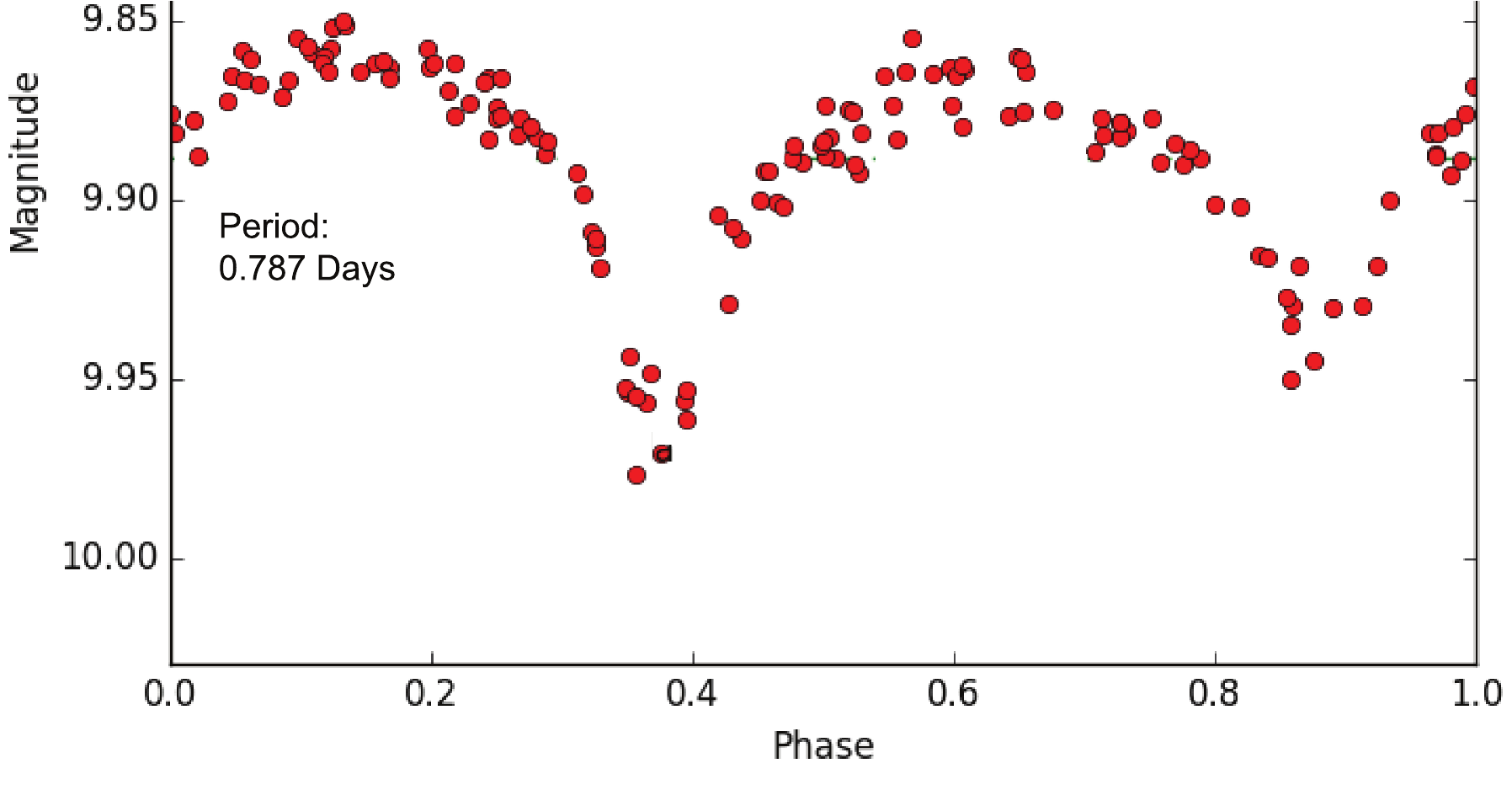}
\caption{Phase-folded light curves for our five eclipsing binary
  detections (all season 1 detections).  {\it Top Left}: TYC
  6799-309-1, {\it Top Right}: TYC 7807-358-1, {\it Middle Left}: HD
  134518, {\it Middle Right}: TYC 7322-822-1, {\it Bottom}: TYC
  7340-720-1.}
\label{EB-Phase}
\end{figure*}

\begin{deluxetable}{rrrrrrr}[htb!]
%\centering
\tabletypesize{\scriptsize}
\setlength{\tabcolsep}{0.03in}
\tablewidth{0pt}
\tablecaption{Periodogram Analysis Results}
\tablehead{
\multicolumn{1}{c}{1SWASP}       & 
\multicolumn{1}{c}{Season Start} & 
\multicolumn{1}{c}{Season End}   & 
\multicolumn{1}{c}{Period}       & 
\multicolumn{1}{c}{Amplitude}\\
\multicolumn{1}{c}{(J)}          & 
\multicolumn{1}{c}{(HJD)}        & 
\multicolumn{1}{c}{(HJD)}        & 
\multicolumn{1}{c}{(Days)}       &
\multicolumn{1}{c}{(Mag)}}
\startdata
111327.46-452332.7 & 2453860.2218 & 2453924.2104 & 0.628 & 0.038\\
111327.46-452332.7 & 2454105.5074 & 2454307.2128 & 0.628 & 0.035\\
111327.46-452332.7 & 2454467.4436 & 2454614.2497 & 0.628 & 0.039\\
111434.43-441824.1 & 2453860.2219 & 2453924.2103 & 0.974 & 0.067\\
\enddata
%\centering
%
\tablecomments{This table contains all of the rotation period
  information extracted from the complete 1689 star sample. The
  remainder of the table is available electronically. Each entry is
  for one SWASP object for a single season between the reported HJD
  dates. For each season, we report the identifiers, season HJDs,
  strongest period(s) in that season, and the amplitude of the
  phase-folded light curve.}
  %\label{Tab:Photometry}
\end{deluxetable}

  %%% Table 1: Periodogram Analysis Giant Table (Small Sample)
\label{Tab:Photometry}
% Full table provided in table1.tex

%\newpage
%\LongTables
%\floattable
\begin{deluxetable}{rrrrrrrrrrrr}[htb!]
\tabletypesize{\scriptsize}
%\setlength{\tabcolsep}{0.001in}
%\tablewidth{0pt}
\textwidth=60pt
\hfuzz=0.64pt
\tablecaption{Data for Sco-Cen Members}
\tablehead{
{1SWASP}   & 
{2MASS}    & 
{Subgroup} & 
{Period}   & 
{SpT}      & 
{Ref.}     &
{$V$}      & 
{$K_S$}    & 
{$A_V$}    & 
{($V$ - $K_S$)$_0$} & 
{\teff}    &
{$M$/$M_\odot$}\\
{(J)}    &
{(J)}    &
{...}    &
{(days)} &
{...}    &
{...}    &
{(mag)}  & 
{(mag)}  & 
{(mag)}  &
{(mag)}  &
{(K)}    &
{...}}
\startdata
111327.46-452332.7 & 11132622-4523427 & LCC & 0.628 $\pm$ 0.002$^a$ & M0.5 & 1 & 12.83 & 8.50 & 0.278 & 4.090 & 3700 & 0.56\\
112102.95-462630.9 & 11210295-4626308 & LCC & 3.798 $\pm$ 0.844$^a$ & --  &  --  & 11.09 & 8.96 & 0.194$^c$ & 1.935 & -- & --\\
114445.40-455106.2 & 11444540-4551062 & LCC & 2.166 $\pm$ 0.016     & -- &  --  & 9.94 & 8.57 & 0.194$^c$ & 1.174 & -- & --\\
124419.32-452523.3 & 12441932-4525235 & LCC & 3.122 $\pm$ 0.044 &  F9V &  3  & 9.60 & 8.00 & 0.279 & 1.350 & 6090 & 1.35\\
\enddata
\tablecomments{
This table contains the members of Sco-Cen reported based on the
criteria described in \S\ref{pgrams}. The remainder of the table is
available electronically. Provided are cross-identifications of SWASP
and 2MASS identifiers, periods and uncertainties, the spectral type
and reference for the 157 stars with spectral types, APASS $V$
magnitudes, 2MASS \vks\ magnitudes, adopted reddening
coefficients, \teff, and mass.\\
($^a$) Stars which have statistically similar periods to the nearest spatial entry in the Variable Star Index \citep{VSX}.\\
($^b$) Stars which have statistically different periods to the nearest spatial entry in the Variable Star Index \citep{VSX}.\\
($^c$) Stars without spectral types use the Sco-Cen sub-group-averaged reddening coefficients for color estimation only.\\
($^d$) The only star in our sample found in \citet{David14}. We find the same period.\\
($^e$) Stars for which APASS $V$ band was unavailable; SPM4 photometry was adopted in its place.\\
References: (1) \citet{Riaz06}, (2) \citet{Pecaut16},
(3) \citet{Houk78}, (4) \citet{Mamajek02}, (5) \citet{Spencer36},
(6) \citet{Torres06}, (7) \citet{Houk82}, (8) \citet{Wichmann97},
(9) \citet{Krautter97}, (10) \citet{Kohler00},
(11) \citet{Preibisch98}, (12) \citet{Houk88}, (13) \citet{Rizzuto15},
and (14) \citet{Luhman12}.}
%\label{Tab:Results}
\end{deluxetable}
  %%% Table 2: Member Results
\label{Tab:Results}
% Full table provided in table2.tex

%\LongTables
\begin{deluxetable}{llcclll}[htb!]
\tabletypesize{\scriptsize}
\setlength{\tabcolsep}{0.03in}
\tablewidth{0pt}
\tablecaption{Data for Other Stars\label{tab:other}}
\tablehead{
\multicolumn{1}{c}{Name}     & 
\multicolumn{1}{c}{Group}    & 
\multicolumn{1}{c}{1SWASP}   & 
\multicolumn{1}{c}{2MASS}    & 
\multicolumn{1}{c}{Period}   & 
\multicolumn{1}{c}{Spectral Types} & 
\multicolumn{1}{c}{Ref.} \\
& & \multicolumn{1}{c}{(J)}  &
\multicolumn{1}{c}{(J)}      &
\multicolumn{1}{c}{(days)} 
}
\startdata
CD-43 6891&...&111434.43-441824.1&11143442-4418240&0.973 $\pm$ 0.004$^b$&K2IV &5\\ 
CD-43 6891&...&111434.43-441824.1&11143442-4418240&39.92 $\pm$ 2.10$^a$ &K2IV &5\\ 
TWA 12          & TWA      & 112105.46-384516.5 &  11210549-3845163  &  3.311 $\pm$ 0.051$^a$  &  M1Ve   &  1\\
TWA 20 & TWA & 123138.06-455859.3& 12313807-4558593& 1.822 $\pm$ 0.014$^a$ &M3IVe&5\\
RX J1539.7-3450 & Lup I    & 153946.41-345102.2 &  15394637-3451027  &  7.127 $\pm$ 0.204$^a$  &  K4     &  2\\
HT Lup          & Lup I    & 154512.86-341730.5 &  15451286-3417305  &  4.304 $\pm$ 0.109$^a$  &  K3Ve   &  1\\
%KW Lup         & foregrd  & 154547.59-302055.7 &  15454761-3020555  &  2.750  &  K2V    &  1$^a$\\
HD 140655       & Lup I    & 154558.54-341341.2 &  15455855-3413411  &  2.753 $\pm$ 0.035  &  F8V    &  3\\
RX J1546.8-3459 & Lup I    & 154645.09-345947.3 &  15464506-3459473  &  2.278 $\pm$ 0.025  &  M0     &  2\\
RX J1548.1-3452 & Lup I    & 154808.93-345253.2 &  15480893-3452531  &  1.423 $\pm$ 0.020  &  M2.5   &  2\\
RX J1548.9-3513 & Lup I    & 154854.16-351318.5 &  15485411-3513186  &  0.933 $\pm$ 0.003  &  K6     &  2\\
IM Lup          & Lup II   & 155609.20-375605.9 &  15560921-3756057  &  7.309 $\pm$ 0.183  &  M0     &  4\\
MML 78          & Lup III  & 160545.00-390606.5 &  16054499-3906065  &  1.261 $\pm$ 0.007$^a$  &  G7V    &  5\\
RX J1607.2-3839 & Lup III  & 160713.73-383923.3 &  16071370-3839238  &  2.418 $\pm$ 0.026  &  K7     &  2\\
IRAS 16051-3820 & Lup III  & 160830.70-382826.8 &  16083070-3828268  &  6.244 $\pm$ 0.129  &  K0?    &  6\\
RX J1608.9-3905 & Lup III  & 160854.27-390605.6 &  16085427-3906057  &  2.005 $\pm$ 0.028$^a$  &  K2     &  2\\
V1095 Sco       & Lup III  & 160939.52-385506.8 &  16093953-3855070  &  2.912 $\pm$ 0.037$^a$  &  K5     &  2\\
Sz 122          & Lup III  & 161016.43-390805.0 &  16101642-3908050  &  0.287 $\pm$ 0.001  &  M2e    &  4\\
RX J1612.0-3840 & Lup III  & 161201.38-384027.5 &  16120140-3840276  &  2.813 $\pm$ 0.033$^a$  &  K5     &  2\\
RX J1620.7-2348 & Oph      & 162045.96-234820.2 &  16204596-2348208  &  3.355 $\pm$ 0.139  &  K3e    &  7\\
RX J1621.0-2352 & Oph      & 162057.86-235234.4 &  16205787-2352343  &  2.097 $\pm$ 0.046$^a$  &  G9IV   &  1\\
EM$^*$ StHa 126& Oph      & 162307.84-230059.8 &  16230783-2300596  &  4.069 $\pm$ 0.235$^a$  &  K2   &  7\\
EM$^*$ SR 6& Oph      & 162528.63-234626.1 &  16252863-2346265  &  3.760 $\pm$ 0.119  &  K2IV   &  5\\
CD-27 10938& Oph      & 162627.35-275651.0 &  16262736-2756508  &  2.065 $\pm$ 0.281$^a$  &  K4IVe   &  5\\
VSS II-28       & Oph      & 162652.81-234312.6 &  16265280-2343127  &  5.595 $\pm$ 0.322$^a$  &  K1IVe  &  5\\
HBC 644         & Oph      & 163104.44-240435.8 &  16310436-2404330  &  0.973 $\pm$ 0.019$^b$  &  K4IVe  &  5\\
V940 Sco        & Oph      & 163201.59-253025.7 &  16320160-2530253  &  2.452 $\pm$ 0.035$^a$  &  K5IVe  &  5\\
V709 CrA        & CrA      & 190134.84-370056.6 &  19013485-3700565  &  2.244 $\pm$ 0.021  &  K2.5   &  8\\
RX J1917.4-3756 & CrA      & 191723.82-375650.3 &  19172382-3756504  &  3.375 $\pm$ 0.045$^a$  &  K0IVe  &  1\\
\enddata
\tablecomments{
($^a$) Stars which have statistically similar periods to the nearest spatial entry in the Variable Star Index \citep{VSX}.\\
($^b$) Stars which have statistically different periods to the nearest spatial entry in the Variable Star Index \citep{VSX}.\\
References: (1) \citet{Torres06}, (2) \citet{Krautter97},
(3) \citet{Houk82}, (4) \citet{Hughes94}, (5) \citet{Pecaut16}
(6) \citet{Antoniucci14} (estimates \teff\ = 5000\,K, which would be
roughly consistent with a K0 \prems\, star on the spectral type
vs. \teff\ scale of \citet{Pecaut13}), (7) \citet{Preibisch98},
(8) \citet{Meyer09}.
%\label{Tab:Other_Stars}
}
\end{deluxetable}
        %%% Table 3: Non-member Results
\label{Tab:Other_Stars}

%\LongTables
\begin{deluxetable}{rrrrrrrr}[htb!]
\tabletypesize{\scriptsize}
%\setlength{\tabcolsep}{0.001in}
%\tablewidth{0pt}
\tablecaption{Adopted Photometry For Stellar Parameters}
\tablehead{
\multicolumn{1}{c}{2MASS} & 
\multicolumn{1}{c}{$B$}   & 
\multicolumn{1}{c}{Ref.}  & 
\multicolumn{1}{c}{$V$}   & 
\multicolumn{1}{c}{Ref.}  & 
\multicolumn{1}{c}{$J$}   & 
\multicolumn{1}{c}{$H$}   & 
\multicolumn{1}{c}{$K_S$} \\
\multicolumn{1}{c}{(J)}   & 
\multicolumn{1}{c}{(mag)} & & 
\multicolumn{1}{c}{(mag)} & & 
\multicolumn{1}{c}{(mag)} & 
\multicolumn{1}{c}{(mag)} & 
\multicolumn{1}{c}{(mag)} }
\startdata
11132622-4523427 & 14.370$\pm$0.012 & DR9 & 12.833$\pm$0.010 & DR9 &  9.415$\pm$0.028 &  8.727$\pm$0.040 &  8.495$\pm$0.031 \\ 
12441932-4525235 & 10.087$\pm$0.035 & H00 &  9.500$\pm$0.022 & H00 &  8.371$\pm$0.027 &  8.072$\pm$0.047 &  7.999$\pm$0.033 \\ 
12480778-4439167 & 10.650$\pm$0.014 & T06 &  9.830$\pm$0.010 & T06 &  8.131$\pm$0.021 &  7.672$\pm$0.055 &  7.513$\pm$0.024 \\ 
12543141-4607361 & 10.250$\pm$0.038 & H00 &  9.713$\pm$0.024 & H00 &  8.376$\pm$0.029 &  7.983$\pm$0.024 &  7.910$\pm$0.023\\
\enddata
\tablecomments{A sample of the adopted photometry for the stars with spectral types.
The remainder of this table is available electronically. AAVSO
Photometric All-Sky Survey (APASS) DR6, DR7 photometry retrieved
from \url{https://www.aavso.org/download-apass-data};
\citet{Henden16}. \\
References -- (DR6) APASS DR6; (DR7) APASS DR7; (DR9) APASS DR9; (P97)
Hipparcos, \citet{Perryman97}; (H00) Converted from Tycho-2 $B_T$,
$V_T$, \citet{Hog00}; (T06) \citet{Torres06}.
%\label{Tab:Pre-AnalysisShort}
}
\end{deluxetable}
  %%% Table 4:
\label{Tab:Pre-AnalysisShort}
% Full table provided in table4.tex

\end{document}